\documentclass[11pt,a4paper]{article}
\usepackage{amssymb}
\usepackage{slashed}
\usepackage{dsfont}
\usepackage{mathrsfs}
\usepackage{chet}
\newcommand{\be}{\begin{equation}}
\newcommand{\ee}{\end{equation}}
\newcommand{\nn}{\nonumber}

\newcommand{\D}{{\cal D}}
\newcommand{\I}{{\cal I}}
\newcommand{\E}{{\mathcal E}}

\newcommand{\cL}{{\cal L}}

\newcommand{\N}{{\cal N}}

\newcommand{\R}{{\cal R}}

\renewcommand{\O}{{\cal O}}

\newcommand{\X}{{\mathcal X}}
\newcommand{\Y}{{\mathcal Y}}

\newcommand{\rmd}{{\rm d}}
\newcommand{\ts}{\textstyle}

\newcommand{\half}{{\tfrac{1}{2}}}
\newcommand{\quar}{{\tfrac{1}{4}}}
\newcommand{\pr}{\partial}

\newcommand{\si}{\sigma}

\newcommand{\vep}{\varepsilon}
\newcommand{\vphi}{\varphi}

\newcommand{\tr}{{\rm tr}}
\newcommand{\Tr}{{\rm Tr}}
\newcommand{\Det}{{\rm Det}}

\newcommand \tphi{\tilde \varphi}
\newcommand \gl{\lambda}
\newcommand\rO{{\rm O}}
\newcommand \ollr{{\raise 8pt\hbox{$\leftrightarrow  \! \! \! \! \! \!$}}}

\newenvironment{acknowledgements}{\vspace{12pt}\begin{center}\textbf{Acknowledgements}\end{center}\vspace{-12pt}}{}
\renewcommand{\ack}[1]{\begin{samepage}\begin{acknowledgements} {#1}
\end{acknowledgements}\end{samepage}}

\makeatletter
  \collectsep{\newline}
\makeatother

\preprint{DAMTP 2015/1}

\title{Structures on the Conformal Manifold\\\vspace{6pt}
in Six Dimensional Theories}

\author{Hugh Osborn$^{a}$ and Andreas Stergiou$^{b}$
\emails{\href{mailto:ho@damtp.cam.ac.uk}{ho@damtp.cam.ac.uk},
\href{mailto:andreas.stergiou@yale.edu}{andreas.stergiou@yale.edu}}}

\affiliation{$^a$Department of Applied Mathematics and Theoretical
Physics, Wilberforce Road,\\ Cambridge CB3 0WA, England\\
\vspace{3pt}
$^b$Department of Physics, Yale University, New Haven, CT 06520
USA}

\abstract{The tensors which may be defined on the conformal manifold for
six dimensional CFTs with exactly marginal operators are analysed by
considering the response to a Weyl rescaling of the metric in the presence
of local couplings. It is shown that there are three symmetric two index
tensors only one of which satisfies any positivity conditions. The general
results are specialised to the six dimensional conformal theory defined by
free two-forms and also to the interacting scalar $\phi^3$ theory at two
loops which preserves conformal invariance to this order.  All three two
index tensor contributions are present.}

\date{January 2015}

\begin{document}

\maketitle

\newsec{Introduction}
If conformal field theories have exactly marginal operators there is a
conformal manifold parameterised by the couplings for the marginal
operators.  In two and four dimensions CFTs with associated conformal
manifolds are not uncommon, at least with $\N=1$ supersymmetry
\cite{Seiberg}. The situation is much less clear in higher dimensions;
whether any non trivial CFTs with marginal operators exist in six
dimensions remains doubtful but not inconceivable \cite{Henn}.  Here we aim
to extend some results obtained in two and four dimensions to the
significantly more complicated case of six.

To this end we consider the response of a CFT extended to a curved space
background to a Weyl rescaling of the metric $\gamma_{\mu\nu}$. In general,
CFTs are invariant under Weyl rescalings of the background metric,
$\gamma_{\mu\nu} \to e^{2\hspace{1pt}\sigma} \gamma_{\mu\nu}$,  up to a
finite sum of local contributions formed from curvature tensors and
$\sigma$, with coefficients commonly referred to as central charges. In two
dimensions there is just the Virasoro central charge $c$, so that the trace
of the energy momentum tensor is proportional to $c\, R$, with $R$ the
scalar curvature which is equal to the two dimensional Euler density $E_2$.
In four dimensions there are just two coefficients $c,a$, which are related
to the square of the Weyl tensor and the four dimensional Euler density
$E_4$.  These results for CFTs on curved backgrounds may be used to
construct effective field theories for a dilaton $\tau$, with terms
$\rO(\tau^2)$ in two dimensions, and $\rO(\tau^3,\tau^4)$ in four
dimensions, which survive on reduction to flat space and are proportional
to $c,a$ respectively. By considering dilaton scattering in four dimensions
the crucial positivity constraints allowing arguments for an irreversible
RG flow between UV and IR fixed points have been obtained \cite{Kom, Luty}.

For CFTs with a conformal manifold it is convenient to allow the couplings
$g^I$ for the marginal operators to be local or $x$-dependent. The
couplings can then be treated as sources for the marginal operators. In
that case there are additional local contributions under a Weyl rescaling
depending on derivatives of $g^I$.  Such terms are restricted by power
counting. In two dimensions this procedure generates  a unique two index
tensor $g_{IJ}$ on the conformal manifold, while in four dimensions a four
index tensor is present also. In two dimensions $g_{IJ}$  is identical with
the metric defined by Zamolodchikov \cite{Zam} in terms of the two point
functions for the scalar operators coupled to $g^I$ and which for unitary
theories is necessarily positive. A similar result applies in the four
dimensional case so the corresponding metric is again positive.

Away from a conformal critical point the response to Weyl rescalings with
local couplings must satisfy Wess--Zumino consistency conditions stemming
from the fact that the Weyl group is Abelian.  The resulting equations
relate the RG flow of the central charge $c$ in two and $a$ in four
dimensions to the corresponding $g_{IJ}$. For positive $g_{IJ}$ the  RG
flow is irreversible \cite{Weyl, JackO, Baume}. In two dimensions this
approach is equivalent to the Zamolodchikov $c$-theorem. In four dimensions
the metric is necessarily positive in the neighbourhood of a fixed point,
but unlike two dimensions there is no simple general non perturbative
argument, although arguments based on dilaton effective actions can be
applied \cite{Baume}.  For renormalisable quantum field theories in four
dimensions the metric and related quantities may be calculated
perturbatively in terms of the vacuum amplitude, most directly with a
curved space background and using local couplings at two loops
\cite{analogs}, but also just restricting to flat space at three loops
\cite{Fortin, FortinTwo, JackO}.

It is natural to consider extensions to higher dimensions, in particular
six.  The dilaton effective action was constructed in \cite{Elvang} and
also \cite{Baume2, Coriano}. The local RG approach was also extended to six
dimensions in \cite{Grinstein}. The number of contributions which it is
necessary to consider increases significantly; in the approach followed in
\cite{Grinstein} there are $\rO(100)$ different consistency conditions to
be analysed. Due to complications arising from the analytic structure of
$3\to 3$ amplitudes there is no derivation of irreversibility of RG flow
along the same lines as that applied in four dimensions \cite{Elvang}, and
recently a two loop calculation in six dimensional $\phi^3$ theory showed
that the metric relevant for RG flow was not positive in this theory
\cite{asix}.

In this paper we endeavour to understand further the complications arising
in six dimensions by considering a six dimensional conformal field theory
with exactly marginal operators. The approach followed here, based on
assuming local couplings for all marginal operators and considering the
response to Weyl rescalings of the metric, defines various tensors on any
conformal manifold. An infinitesimal Weyl rescaling determines the trace of
the energy momentum tensor. As is well known, in six dimensions on a curved
background with fixed couplings and neglecting scheme dependent
contributions, this is expressible in terms of three scale dimension six
Weyl invariants, with coefficients $c_1,c_2,c_3$, and the topological Euler
density $E_6$, with coefficient $a$ \cite{Bonora, Deser}. Thus
$c_1,c_2,c_3,a$ may be regarded as the central charges in six dimensions,
corresponding to the two dimensional $c$ and four dimensional $c,a$.  With
local couplings to marginal operators it is further possible to obtain
three rank two symmetric tensors, as well as rank four and rank six
tensors. One rank two symmetric tensor can be related to the two point
function for marginal operators and is therefore positive.  This may then
be taken as a metric for the conformal manifold. However, contrary to the
case in two and four dimensions, this is not the tensor that features in
the equation for the RG flow of $a$. The additional symmetric tensors
present in six dimensions are constructed in terms of the Weyl tensor and
so are absent in any conformally flat space.

In the next section we review the response of a CFT containing exactly
marginal operators in four dimensions and then consider the extension to
six.  In six dimensions it is necessary to consider Weyl transformations
which are rather more involved than in four. Besides the Weyl tensor the
results can be expressed more simply in terms a basis involving the Cotton
and Bach tensors \cite{Feff}.  Their definitions and some basic properties
are reviewed in appendix \ref{conformal}. It is also necessary to consider
various conformally covariant differential operators  which extend the
conformal Laplacian $\Delta_2 = - \nabla^2 + \xi R$, where $\xi =
(d-2)/4(d-1)$ with $d$ the spacetime dimension. In four dimensions the
results involve $\Delta_4$, the conformal extension of $(\nabla^2)^2$,
while in  six dimensions it is necessary to consider the Branson operator
$\Delta_6$ \cite{Branson} whose leading term is $-(\nabla^2)^3$.

As an illustration of these results we consider in section \ref{TwoForms}
the conformal theory in six dimensions which is obtained from the quantum
field theory of free two-forms. In this case we may introduce a local
coupling in the action as $1/g^2$ which acts as a source for the dimension
six scalar operator formed by the gauge invariant classical Lagrangian
density. After suitable gauge fixing we determine the one loop anomalous
contributions under a Weyl rescaling of the metric, extending the results
in \cite{Bastianelli} to include contributions involving derivatives of
$g$. The results fit the general structure determined in section
\ref{response}.

In section \ref{phicubed} results obtained from calculations at two loops
for $\phi^3$ theory on a curved background with local couplings are also
presented.  This theory has non zero $\beta$-functions and conformal
invariance  is broken but perturbative calculations should satisfy the
constraints obtained in section \ref{response} to lowest order. We also
present results for the central charges $c_1,c_2,c_3,a$ to $\rO(g^2)$. To
ensure that the results are compatible with the general analysis it is
necessary to ensure when using dimensional regularisation that the one loop
counterterms are such as to ensure the initial free theory is conformal
away from $d=6$.  Although $\phi^3$ theory is problematic, since it lacks a
minimum energy ground state, we assume it may be stabilised by a small
$\phi^4$ term and that it may then still be used to define an effective
conformal theory, at least to leading order.

We also consider in section \ref{positivity} some positivity conditions
which are obtained by relations to two point functions. These serve as a
check on the results for $c_3$ which is related to the energy momentum
tensor two point function and also a two index tensor on the space of
marginal couplings which is related to the two point function for the
exactly marginal dimension six scalar operators. The coefficients $c_1,c_2$
as well as $c_3$ determine the energy momentum tensor three point function.
This also satisfies positivity restrictions related to the energy flux at
infinity \cite{Hofman} and these are shown to  be satisfied to lowest order
beyond free theory by $\phi^3$ theory.

Various details are contained in four appendices. In appendix
\ref{conformal} we present a detailed summary of results for conformal
tensors, the Weyl, Cotton and Bach tensors, and also differential operators
which transform nicely under Weyl rescaling of the metric and  are relevant
for our calculations. We also give an expression for the coincident limit
of the Seeley--DeWitt coefficient $a_3$, which determines the one loop
results, in terms of the basis of conformal tensors. In appendix
\ref{sixddilaton} we describe briefly the six dimensional results obtained
by integrating the infinitesimal Weyl rescaling of the metric. Appendices
\ref{resFermions} and \ref{resTwoForms} contain the detailed results
necessary to calculate the coincident limit of $a_3$ for fermions and
two-forms respectively.

\newsec{Response to Weyl Rescalings for CFTs}[response]
In general the vacuum functional $W$, depending on the metric and couplings,
for a CFT responds to an infinitesimal Weyl rescaling,
$\delta_\sigma  \gamma_{\mu\nu}= 2 \hspace{1pt}\sigma  \hspace{1pt}\gamma_{\mu\nu}$, in even $d$ dimensions according to
\be
(4\pi)^{\frac{d}{2}} \, \delta_{\sigma} W = \int \rmd^d x \sqrt{-\gamma}\;
\sigma L_d \, ,
\label{basic}
\ee
with $L_d$ a local scalar of dimension $d$ formed from the metric, the couplings and derivatives.
In general $L_d$ is constrained by the integrability conditions following from
$(\delta_\sigma \delta_{\sigma'}  - \delta_{\sigma'} \delta_\sigma)W =0$. We initially consider solutions such that
\be
\delta_\sigma L_d + d \, \sigma L_d = \nabla_\mu ( X_d\vphantom{X}^{\mu\nu} \, \pr_\nu \sigma) \, ,
\qquad X_d\vphantom{X}^{\mu\nu} = X_d\vphantom{X}^{\nu\mu} \, .
\label{consis}
\ee
We assume that in \eqref{basic} $L_d$ has the freedom
\be
L_d \sim L_d + \nabla_\mu \nabla_\nu Z_d\vphantom{Z}^{\mu\nu}\, ,
\label{freedom}
\ee
since such contributions can in general be cancelled by local contributions to $W$.
For variations \eqref{freedom} compatible with \eqref{consis} then
\be
X_d\vphantom{X}^{\mu\nu} \sim  X_d\vphantom{X}^{\mu\nu} + 2\,  Z_d\vphantom{Z}^{\mu\nu} - \gamma^{\mu\nu} \,
Z_d\vphantom{Z}^{\lambda}\vphantom{Z}_\lambda \quad \mbox{if} \quad \delta_\sigma
Z_d\vphantom{Z}^{\mu\nu}+ d\, \sigma Z_d\vphantom{Z}^{\mu\nu} = 0 \, .
\ee

Under a finite rescaling \eqref{basic} extends to
\be
(4\pi)^{\frac{d}{2}} \big ( W\big [e^{2\hspace{1pt}\sigma} \gamma_{\mu\nu} \big ] -
W\big [\gamma_{\mu\nu} \big ] \big )
=  \int \rmd^d x \sqrt{-\gamma}\; \cL_d(\sigma) \, ,
\label{wsig}
\ee
where $\cL_d(\sigma) $ is obtained by a Taylor expansion,
\eqna{\cL_d(\sigma) = {}& \sigma L_d - \pr_\mu \sigma  \pr_\nu \sigma \,
{\textstyle \sum_{r\ge 0}} \tfrac{1}{(r+2)!} \,
X_{d,r}{\vphantom{X}}^{\mu\nu} + \nabla_\mu J^\mu  \, , \\
X_{d,r+1}{\vphantom{X}}^{\mu\nu}  ={}&  (\delta_\sigma + d \, \sigma)
X_{d,r}{\vphantom{X}}^{\mu\nu} \, , \qquad
X_{d,0}{\vphantom{X}}^{\mu\nu}  = X_{d}{\vphantom{X}}^{\mu\nu}  \, ,}[Xrecur]
so that $X_{d,r}{}^{\mu\nu} = \rO(\sigma^r)$ and $J^\mu$ is
arbitrary. The sum in \eqref{Xrecur} truncates after a finite number of terms.

 Before proceeding to the six dimensional case we recapitulate previous results
 obtained in four dimensions \cite{JackO}. The extra terms involving derivatives of
 the couplings depend on a symmetric two index tensor $g_{IJ}$ and also a four index tensor
 $c_{IJKL}$. It is natural to express the contributions to $L_4$ using the Christoffel
 connection formed from $g_{IJ}$,
 \be
 \Gamma^{I}{\!}_{JK} = \half\,  g^{IL}  ( \pr_J g_{LK} + \pr_K g_{LJ} - \pr_L g_{JK}) \, ,
\qquad g^{IJ} = (g^{-1})^{IJ} \, .
 \label{Christ}
 \ee
We may also allow for a background gauge field $A_\mu \in \mathfrak{g}$ coupled to
conserved currents.  If $F_{\mu\nu}$ is the associated field strength, then
\eqna{L_4 = {}&  c\, W^{\rho\mu\nu\lambda}W_{\rho\mu\nu\lambda} - a\, E_4
- \quar \, \kappa_{ab} \, F_a{\vphantom{F}} ^{\mu\nu}  F_{b\, \mu\nu}  \\
&{}+  \half \, g_{IJ} \, D^2 g^I  D^2 g^J -
g_{IJ} \, \pr^\mu  g^I (2P_{\mu\nu} - \gamma_{\mu\nu} {\hat R}) \,
\pr^\nu g^J \\
&{} + \half \, c_{IJKL} \,
\pr^\mu g^I \pr_\mu g^J  \,\pr^\nu g^K \pr_\nu g^L \, .}[Lfour]
Here,
\be
D^2 g^I  = \nabla^2 g^I + \Gamma^I{\!}_{JK}  \pr^\mu g^J \pr_\mu g^K \, ,
\label{Dsq}
\ee
$E_4$ is the Euler density, given by \eqref{Efour} for $d=4$,
and $P_{\mu\nu}$ and $\hat{R}$ are the Schouten tensor and its trace
given by \eqref{defPR} for $d=4$. In \eqref{Lfour} clearly $g_{IJ}=g_{JI}, \, c_{IJKL}
= c_{(IJ)(KL)} = c_{KLIJ}$ and $\kappa_{ab} $ is a symmetric invariant bilinear form
in a convenient basis $\{t_a\}$ for ${\mathfrak g}$ so that, for any $X\in {\mathfrak g}$,
$X  = X_a t_a $. If ${\mathfrak g}$ is simple then $\kappa_{ab} \to \kappa \, \delta_{ab}$.
We may also extend $\pr_\mu g^I \to \pr_\mu g^I + A_{a\, \mu}(T_a g)^I$ but for
simplicity we neglect such contributions here.

It is straightforward to check that \eqref{Lfour} satisfies \eqref{consis}
with
\be
X_4\vphantom{X}^{\mu\nu} = - 8\hspace{1pt}a \, G_4\vphantom{G}^{\mu\nu} + g_{IJ} \big ( 2 \,  \pr^\mu g^I  \pr^\nu g^J -
\gamma^{\mu\nu}  \, \pr^\lambda g^I  \pr_\lambda g^J  \big ) \, ,
\label{Xfour}
\ee
and $G_4\vphantom{G}^{\mu\nu} $ as in \eqref{Gfour} with $d=4$, so long as $a$ is constant.
From \eqref{Xrecur} it is easy to see that
\be
X_{4,1}{}^{\mu\nu} = 16\hspace{1pt}a ( \nabla^\mu \pr^\nu \sigma -
\gamma^{\mu\nu} \, \nabla^2 \sigma ) \, , \qquad
X_{4,2}{}^{\mu\nu} =  -16\hspace{1pt} a ( 2\, \pr^\mu \sigma \pr^\nu \sigma
+ \gamma^{\mu\nu}\,  \pr^\lambda \sigma \pr_\lambda \sigma ) \, .
\label{Xfour1}
\ee
Using \eqref{Xfour}, \eqref{Xfour1} in \eqref{Xrecur} gives
\eqna{
\cL_4(\sigma) = {}& \sigma \, L_4 + \half \,  g_{IJ} \big ( 2 \,  \pr^\mu g^I  \pr^\nu g^J -
\gamma^{\mu\nu}  \, \pr^\lambda g^I  \pr_\lambda g^J  \big ) \pr_\mu \sigma \pr_\nu\sigma\\
&{} + 4\hspace{1pt} a \big (G_4\vphantom{G}^{\mu\nu} \pr_\mu \sigma \pr_\nu\sigma + \nabla^2 \sigma \,
\pr^\mu \sigma \pr_\mu\sigma + \half \, (\pr^\mu \sigma \pr_\mu\sigma )^2
\big ) \, ,}[]
 which reproduces the well known results for the four dimensional
dilaton effective action and the $\pr g$ terms calculated in \cite{JackO}.

In  six dimensions we follow a similar route by determining the general form for
$L_6$ satisfying
\eqref{consis}. There are various contributions which may be analysed independently.
For any six dimensional CFT in the absence of local couplings $L_6$ is given by just
\be
L_{6}^R = {\ts \sum_{i=1,2,3}}\,  c_i \, I_i + a \, E_6 \, ,
\label{Lsix}
\ee
where an appropriate basis for the dimension six conformal scalars $I_i$, and
also an explicit expression for the Euler  density $E_6$, are given in
appendix \ref{conformal}. $I_1,I_2$ are the two independent scalars cubic in the
Weyl tensor while $I_3= W^{\rho\mu\nu\lambda} \nabla^2
W_{\rho\mu\nu\lambda} + \cdots$.
Since $\delta_\sigma I_i + 6\hspace{1pt} \sigma I_i =0 $ and
$\delta_\sigma E_6 + 6\hspace{1pt} \sigma E_6 =
24\, \nabla_\mu ( G_6\vphantom{G}^{\mu\nu} \pr_\nu \sigma)$, with $G_6\vphantom{G}^{\mu\nu}$ the six dimensional
generalisation of the Einstein tensor, it is easy to verify that \eqref{Lsix}
satisfies \eqref{consis} with
\be
X_{6}^{R\,\mu\nu} = 24\, a\, G_6\vphantom{G}^{\mu\nu} \, .
\label{Xsix}
\ee

There are also three potential dimension six conformal scalars formed from
$F_{\mu\nu}$ for which we may take
\eqna{
L_6^F = {}& - \quar \, \Big(\kappa_{ab} \big ( F_a{\vphantom{F}}^{\mu\nu}\,
(\D^2 F_{\mu\nu})_b
- 4\, {\hat R} \,  F_a{\vphantom{F}}^{\mu\nu}F_{b\, \mu\nu} \big )
+ 2\,( \nabla_\mu\nabla_\nu + 2 P_{\mu\nu} ) \big (\kappa_{ab} \,
F_a{\vphantom{F}}^{\mu\lambda}F_b{\vphantom{F}}^{\nu}{\!}_\lambda \big
)\Big) \\
&{} - \quar \, {\hat\kappa}_{ab} \, W_{\mu\nu\lambda\rho}\,
F_a{\vphantom{F}}^{\mu\nu}\, F_b{}^{\lambda\rho}
+ \tfrac13 \,  f_{abc}\,  F_a{\vphantom{F}}^{\mu\nu} F_{b\, \nu\lambda}
F_c{}^\lambda{\!}_\mu  \, ,}[LsixF]
with $\kappa_{ab}, {\hat\kappa}_{ab}$ symmetric invariant tensors and
$f_{abc}$ an antisymmetric invariant tensor.

For free scalars, fermions and also two-form gauge fields the coefficients
$c_i, a$ were calculated in \cite{Bastianelli}. For scalars and fermions the results
can be straightforwardly extended to include background  gauge fields.\footnote{A
calculation for the gauge field contributions on flat space backgrounds is given in
\cite{Petkou}. Dropping curvature and total derivative terms our results in a similar basis give
$L_{6,{\rm scalars}} = - \tfrac{1}{60} \, \tr ( \D_\mu F^{\mu\lambda} \D_\nu F^{\nu}{}_\lambda )
+ \tfrac{1}{90} \, \tr ( F^{\mu\nu} F_{ \nu\lambda} F^\lambda{\!}_\mu  ) $ and
$L_{6,{\rm fermions}} = - \tfrac{8}{15} \, \tr ( \D_\mu F^{\mu\lambda} \D_\nu F^{\nu}{}_\lambda )
- \tfrac{4}{45} \, \tr ( F^{\mu\nu} F_{ \nu\lambda} F^\lambda{\!}_\mu  ) $.
}
For $t_a$ the real antisymmetric or
anti-hermitian generators determining the gauge couplings to scalars or fermions,
then, letting  $\kappa_{ab} = - \kappa \, \tr (t_a t_b) , \,
{\hat \kappa}_{ab} = - {\hat \kappa} \, \tr (t_a t_b)$,
$f_{abc} = - f \, \tr(t_{\smash{[a}} t_b t_{\smash{c]}})$,
we have
\be
\begin{tabular}{l*{7}{c}}
\noalign{\vskip -3pt}
& $ 7! c_1$ & $7! c_2$ & $7! c_3$ & $7! a$  &  \ $\kappa$ & \ ${\hat \kappa}$ & \ $f$ \\
\noalign{\vskip 2pt}
\hline
\noalign{\vskip 5pt}
scalars & $- \tfrac{28}{3}$ & $\tfrac{5}{3}$ & $2$ & $\tfrac{5}{9} $ & \ $\tfrac{1}{30}$ &$ \ \tfrac{1}{18}$
& \ $\tfrac{1}{15}$  \\
\noalign{\vskip 5pt}
fermions           & $-\tfrac{896}{3}$ & $-32$ & $40$ & $\tfrac{191}{9}$ &\ $\tfrac{16}{15}$ &
\ $\tfrac{8}{9}$ & \ $\tfrac{52}{15}$  \\
\noalign{\vskip 5pt}
two-forms    &  $-\tfrac{8008}{3}$ & $-\tfrac{2378}{3}$  & $180$  & $442$   \\
\end{tabular}\ .
\label{table}
\ee

Our main motivation in this paper is to consider contributions
depending on derivatives of the couplings.
We first consider terms which are the direct extension of the terms involving $g_{IJ}$ in
 \eqref{Lfour}.  This can be constructed starting from a leading contribution involving
six derivatives
\be
S_1 = \half \, g_{IJ} D^\mu D^2 g^I  D_\mu D^2 g^J
+ \half \,  \R_{IKLJ}  D^2 g^I \pr^\mu g^K \pr_\mu g^L D^2 g^J  \, ,
\label{S1}
\ee
with $D^2 g^I$ defined as in \eqref{Dsq} and
\be
D^\mu D^2 g^I = \pr^\mu  D^2 g^I  + \Gamma^I{\!}_{KL} \pr^\mu g^K D^2 g^L  \, .
\ee
In \eqref{S1}  $\R_{IKLJ}$ is the Riemann
tensor defined as usual in terms of the Christoffel connection
$\Gamma^I{\!}_{KL}$
in \eqref{Christ}.
The Weyl variation of $S_1$ gives
\eqna{\delta_\sigma S_1 + 6\hspace{1pt} \sigma  S_1
= {}& - 8\, g_{IJ} D^2 g^I D^\mu \pr^\nu g^J \nabla_\mu \pr_\nu \sigma
+  3\, g_{IJ} D^2 g^I D^2 g^J \nabla^2 \sigma \\
&{} + 4 \, g_{IJ} \pr^\mu g^I \pr^\nu g^J \big ( \nabla_\mu \pr_\nu \nabla^2
\sigma + 8 \, \nabla_\mu ( P_{\nu\lambda}\, \pr^\lambda \sigma )
+ 2 \, \nabla_\mu (\hat{R}\, \pr_\nu \sigma )\big ) \\
&{} - 2 \, g_{IJ} \pr^\mu g^I \pr_\mu g^J \big ( \nabla^2 \nabla^2
\sigma + 8 \, \nabla^\nu ( P_{\nu\lambda}\, \pr^\lambda \sigma  )
+2 \, \nabla^\nu ( \hat{R}\, \pr_\nu \sigma  )\big ) \\
&{}+ \nabla_\mu \big (4\,  g_{IJ} D^2 g^I D^\mu ( \pr^\nu g^J \pr_\nu \sigma)
- 3 \, g_{IJ} D^2 g^I D^2 g^J \, \pr^\mu \sigma \\
& \hskip 1cm {}- 4 \, g_{IJ}  \pr^\mu g^I \pr^\nu g^J \, (\pr_\nu \nabla^2 \sigma
+ 8 \, P_{\nu\lambda}\,\pr^\lambda \sigma +2 \, \hat{R}\,\pr_\nu \sigma) \\
& \hskip 1cm {}+  2 \, g_{IJ} \pr^\lambda g^I \pr_\lambda g^J \,
( \pr^\mu \nabla^2 \sigma +  8 \, P^{\mu\nu}\,\pr_\nu \sigma + 2\,
\hat{R}\,\pr^\mu \sigma) \big ) \,
.}[]
If we then add the four derivative term
\eqna{S_2 ={}&  - 4 \,  g_{IJ} ( D^2 g^I  D^\mu \pr^\nu g^J
+ D^\mu \pr^\nu  g^I D^2 g^J ) \, P_{\mu\nu}
+ 3 \, g_{IJ} D^2 g^I  D^2 g^J {\hat R} \\
&{} + \nabla_\mu \big ( 4 \, P^{\mu\lambda} g_{IJ} \pr_\lambda g^I D^2 g^J
- 4\,  g_{IJ} \pr^\mu g^I \pr^\nu g^J \pr_\nu {\hat R} + 2\,
g_{IJ} \pr^\lambda g^I \pr_\lambda g^J \pr^\mu {\hat R} \big ) \, ,}[S2]
we may obtain
\be
\delta_\sigma (S_1+S_2) + 6\hspace{1pt} \sigma  (S_1+S_2)  =
g_{IJ} \pr^\mu g^I  \pr^\nu g^J A_{\mu\nu}  +
\nabla_\mu (X_{6}^{g\,\mu\nu}\, \pr_\nu \sigma) \, ,
\label{varS}
\ee
for
\eqna{A_{\mu\nu} = {}& 4 \, \nabla_\mu \pr_\nu \nabla^2 \sigma
+ 32 \, \nabla^\lambda(P_{\lambda(\mu}\pr_{\nu)}\sigma) - 16\,
\nabla^\lambda(P_{\mu\nu}\pr_{\lambda}\sigma)+  16 \, \nabla_{(\mu}(
P_{\nu)\lambda} \pr^\lambda \sigma) \\
&{}- 8 \,  \nabla_{(\mu}({\hat R}\, \pr_{\nu)} \sigma) \\
&{}+ \gamma_{\mu\nu}\big (  - 2 \, \nabla^2 \nabla^2 \sigma - 8\,
\nabla^\mu(P_{\mu\nu} \pr^\nu \sigma)
+ 4 \, \nabla^\mu ({\hat R} \, \pr_\mu \sigma ) \big )}[]
and
\eqna{X_{6}^{g\,\mu\nu}  = {}& 4\, g_{IJ} D^2 g^I D^\mu \pr^\nu g^J - 3 \, \gamma^{\mu\nu}
g_{IJ} D^2 g^I D^2 g^J \\
&{}  + 8 \, P^{\mu\nu} g_{IJ} \pr^\lambda g^I \pr_\lambda g^J
 - 16\big  (  P^{\mu\lambda} g_{IJ} \pr_\lambda g^I \pr^\nu g^J
+  g_{IJ} \pr^\mu g^I \pr_\lambda g^J P^{\lambda\nu} \big )  \\
&{} + 16 \,  g_{IJ} \pr^\mu g^I \pr^\nu g^J {\hat R}
+  \gamma^{\mu\nu} \big ( 16 \,  g_{IJ} \pr^\lambda g^I \pr^\rho g^J P_{\lambda\rho}
- 8\, \pr^\lambda g^I \pr_\lambda  g^J {\hat R} \big )\, ,}[X6g]
where $P_{\mu\nu}, {\hat R}$ are given by \eqref{defPR} with $d=6$.
The remaining $A_{\mu\nu}$ terms in \eqref {varS} may be cancelled by taking
\eqna{S_3 = {}& 4\, g_{IJ} \pr^\mu g^I  \pr^\nu g^J \big (  B_{\mu\nu}
+ 6 \, P_{\mu\lambda}P_\nu{}^\lambda -4 \, P_{\mu\nu}{\hat R} +
\nabla_\mu \pr_\nu {\hat R}  \big ) \\
&{}-2\,   g_{IJ} \pr^\mu g^I  \pr_\mu g^J \big ( 2\, P_{\rho\lambda}
P^{\rho\lambda} -2 \, {\hat R}^2 +  \, \nabla^2 {\hat R} \big ) \, .
}[S3]
Hence, we may satisfy \eqref{consis} for $d=6$ by taking
\eqna{
L_{6}^g = {}& -S_1-S_2-S_3 \\
&{}+  g_{1,IJ} \pr^\mu g^I \pr^\nu g^J \, W_{\mu\lambda\rho\omega}
W_\nu{}^{\lambda\rho\omega} +
 g_{2,IJ} \pr^\mu g^I \pr_\mu g^J \, W_{\nu\lambda\rho\omega}
W^{\nu\lambda\rho\omega} \, .}[L6g]
The terms involving $g_{IJ}$  are a  natural generalisation of the unique $L_{4}^g$,
implicitly defined
by \eqref{Lfour}, and $L_{2}^g = - \half \, g_{IJ} \pr^\mu g^I \pr_\mu g^J$. The sign
is chosen so as to ensure later that $g_{IJ}$ is positive in unitary theories.
In six dimensions there are further possibilities involving rank two tensors
which are formed in terms of the Weyl
tensor, as included in \eqref{L6g}. For these terms \eqref{consis} becomes essentially trivial.

Further contributions to $L_6$ involve at least four $g$'s with derivatives.
To construct these we first consider
\eqna{
T_1 = {}& j_{1,IJKL}  \, \half \big (
\nabla_\rho h^{IJ\,\mu \rho} \, \nabla^\lambda h^{KL}{}_{\mu\lambda}
-  \nabla^\lambda h^{IJ\, \mu \nu} \, \nabla_\lambda h^{KL}{}_{\mu\nu} \big ) \\
&{} + j_{2,IJKL}  \, \half \, \pr^\lambda ( \pr^\mu g^I \pr_\mu g^J ) \;
\pr_\lambda  ( \pr^\nu g^K \pr_\nu g^L ) \\
&{}+ j_{3,IJKL}  \,  \big ( \nabla^\mu \pr^\nu g^I \, \nabla_\mu \pr_\nu g^J -
\quar \, \nabla^2 g^I \, \nabla^2 g^J \big ) \pr^\rho g^K \pr_\rho g^L \, ,
}[T1]
for $h^{IJ}{}_{\mu \nu}$ symmetric and traceless,
\be
h^{IJ}{}_{\mu \nu} = \pr_{(\mu} g^I \pr_{\nu)} g^J - \tfrac{1}{6} \,
\gamma_{\mu\nu} \, \pr^\lambda g^I \pr_\lambda g^J \, ,
\ee
and $j_{i,IJKL} = j_{i,(IJ)(KL)} = j_{i,KLIJ} $.
In this case
\eqna{\delta_\sigma T_1 + 6\hspace{1pt} \sigma T_1 = {}&
\pr_\lambda  \sigma \, j_{1,IJKL} \big (
2 \, \nabla_\mu ( h^{IJ\, \lambda \nu} h^{KL\,\mu }{}_{\nu} )
+  \pr^\lambda ( h^{IJ\, \mu \nu} \,h^{KL}{}_{\mu\nu})  \big ) \\
&{}- \pr_\lambda \sigma \,(  j_{2,IJKL} + j_{3,IJKL} ) \, \pr^\lambda \big (
 \pr^\mu g^I \pr_\mu g^J \,\pr^\nu g^K \pr_\nu g^L \big ) \, .}[]
Terms involving two derivatives of $\sigma$ may be cancelled by
\eqna{
T_2 = {}&-  j_{1,IJKL}  \, \big (
2 \,  P_{\mu\nu}\, h^{IJ\,\mu \lambda} \, h^{KL\, \nu}{}_\lambda  +
{\hat R} \,  h^{IJ\, \mu \nu} \, h^{KL}{}_{\mu\nu} \big ) \\
&{} +(  j_{2,IJKL} +  j_{3,IJKL} )  \, {\hat R} \, \pr^\mu g^I \pr_\mu g^J  \,
\pr^\nu g^K \pr_\nu g^L \\
&{} -    \pr_M j_{1,IJKL} \big ( h^{IJ\, \mu\lambda} \, h^{KL}{}_{\mu \rho}
\,   \nabla^\rho\pr_\lambda g^M
  - \half \, h^{IJ\, \mu\nu}\,   h^{KL}{}_{\mu\nu} \,\nabla^2 g^M \big ) \\
&{} - \quar  \, \pr_M ( j_{2,IJKL} + j_{3,IJKL} ) \,  \pr^\mu g^I \pr_\mu g^J \,
 \pr^\nu g^K \pr_\nu g^L   \, \nabla^2  g^M \, .}[T2]
Hence
\be
L_6^j = T_1 + T_2 + j_{4,IJKL}  \, W_{\mu\lambda\nu \rho} \,
h^{IJ\, \mu \nu} \,h^{KL\, \lambda \rho}
\label{L6j}
\ee
satisfies \eqref{consis} with
\eqna{X_6^{j\,\mu\nu} = {}&   j_{1,IJKL}\big  ( 2 \, h^{IJ\, \mu \lambda}\,
h^{KL\, \nu}{}_\lambda + \gamma^{\mu \nu} \, h^{IJ\, \lambda\rho}
h^{KL}{}_{\lambda\rho} \big ) \\
&{} - \gamma^{\mu\nu} \,(  j_{2,IJKL} + j_{3,IJKL} ) \,
\pr^\lambda g^K \pr_\lambda g^J \, \pr^\rho g^K \pr_\rho g^L \, .}[X6j]
In \eqref{L6j} we have allowed for a possible trivial term involving the Weyl tensor.
If in \eqref{T1} $ \nabla_\lambda h^{KL}{}_{\mu\nu} \to  D_\lambda h^{KL}{}_{\mu\nu}$
and similarly
$\pr_\lambda  ( \pr^\nu g^K \pr_\nu g^L ) \to D_\lambda  ( \pr^\nu g^K \pr_\nu g^L )$,
$ \nabla^\mu \pr^\nu g^I \to D^\mu \pr^\nu g^I$, $\nabla^2 g^I \to D^2 g^I$,
with $D_\lambda$ the covariant derivative including the the Christoffel
connection \eqref{Christ}, then correspondingly in \eqref{T2}
$ \pr_M j_{i,IJKL} \to  D_M j_{i,IJKL}$
with $D_M$ the covariant extension of $\pr_M$.

The remaining potential contribution to $L_6$ involves six $g$'s with
derivatives,
\be
L_6^k = \half \, k_{IJKLMN} \, \pr^\mu g^I \pr_\mu g^J \,
\pr^\nu g^K \pr_\nu g^L \, \pr^\omega g^M \pr_\omega g^N \, ,
\label{X6k}
\ee
defining a rank six tensor with appropriate symmetries.

\newsec{Two-Forms}[TwoForms]

In six dimensions there are three free conformal field theories. In four
dimensions with abelian gauge fields it is still possible to determine the
leading one loop contribution to the metric on the conformal manifold. Here
we describe the analogous calculation in six dimensions following the
approach described in \cite{Local} and extending the six dimensional
results in \cite{Bastianelli}.

For a two-form $B_{\mu\nu} \in \Omega^{(2)}$, where $\Omega^{(n)}$ is the space
of $n$-forms comprised of antisymmetric $n$-index tensors,  the starting Lagrangian is
just\footnote{The exterior derivative $\rmd : \Omega^{(n)} \to \Omega^{(n+1)}$,
is defined so that $(\rmd F)_{\mu_1 \ldots \mu_{n+1} }
= (n+1)\, \pr_{[\mu_1} F_{\mu_2 \ldots \mu_{n+1}]}$ for $F_{\mu_1\ldots \mu_n} \in \Omega^{(n)}$
and is independent of the metric.
The adjoint $\delta : \Omega^{(n)} \to \Omega^{(n-1)}$ is correspondingly given by
$(\delta F)_{\mu_1 \ldots \mu_{n-1}}
= - \frac{1}{\sqrt{-\gamma}} \gamma_{\mu_1\nu_1}
\!\cdots \gamma_{\mu_{n-1}\nu_{n-1}} \pr_{\omega}
( \sqrt{-\gamma}\, \gamma^{\omega\lambda}
\gamma^{\nu_1 \rho_1}\!\cdots \gamma^{\nu_{n-1}
\rho_{n-1}} F_{\lambda \rho_1\ldots \rho_{n-1}})
= - \nabla^\lambda F_{\lambda \mu_1 \ldots \mu_{n-1}}$.
Of course $\rmd^2 = \delta^2=0$.}
\be
{\mathscr L}
= - \frac{1}{12g^2}\, (\rmd B)^{\mu\nu\omega} (\rmd B)_{\mu\nu\omega} \, .
\label{twoL}
\ee
This is invariant under gauge transformations
$B_{\mu\nu} \to B_{\mu\nu} + (\rmd A)_{\mu\nu}$, $A_\mu \in \Omega^{(1)}$.
It is convenient here to add the covariant Feynman gauge fixing term,
\be
{\mathscr L}_{\rm g.f.} = - \frac{g^2}{2}\, \big (\delta (\tfrac{1}{g^2} B)\big ){}^\mu
\big (\delta (\tfrac{1}{g^2}B)\big ){}_\mu \, .
\label{twoGF}
\ee
Rescaling $B_{\mu\nu} \to g B_{\mu\nu}$ the quantum theory is
defined in terms of the functional determinants of the Laplacians
\be
\Delta^{(n)} = \delta' \rmd' + \rmd' \delta' : \Omega^{(n)} \to \Omega^{(n)} \, ,
\qquad \rmd' = \tfrac{1}{g} \, \rmd \, g \, , \quad
\delta' = g \, \delta \, \tfrac{1}{g} \, ,
\label{laplace}
\ee
so that \cite{Bastianelli}
\be
W = - \half  \ln \Det\,  \Delta^{(2)} +  \ln \Det\,  \Delta^{(1)}
-  \tfrac{3}{2}  \ln \Det \, \Delta^{(0)}  \, .
\label{Wtwo}
\ee
$\Delta^{(1)}$ is related to a fermionic vector ghost and $\Delta^{(0)}$
to a bosonic scalar ghost; the degrees of freedom in $d$ dimensions are then
$\half d(d-1) - 2\, d +3  = \half (d-2)(d-3)$.

Continuing to a Euclidean metric the functional determinant of an elliptic
differential operator $\Delta$ may be defined in terms of the heat kernel by
\be
- \ln \Det \, \Delta = \zeta_\Delta\! {}'(0) \, , \qquad
 \zeta_\Delta (s) = \frac{1}{\Gamma(s)} \int_0^\infty \!\!\! \rmd \tau \,
\tau^{s-1}\,  \Tr \big ( e^{-\tau \Delta} \big ) \, .
\label{zeta}
\ee
Under Weyl rescaling of  the metric, for $F_{\mu_1 \dots \mu_{n} }$ an $n$-form,
$\delta_\sigma (\rmd F)_{\mu_1 \dots \mu_{n+1} } = 0$, whereas
$ \delta_\sigma  (\delta F)_{\mu_1 \dots \mu_{n-1}}
=  - 2(d-n+1) \sigma \, (\delta F)_{\mu_1 \dots \mu_{n-1}}
+ 2(d-n) (\delta \, \sigma F)_{\mu_1 \dots \mu_{n-1}}$. Hence, with $d=6$,
\eqna{
\delta_\sigma \, \Delta^{(2)} = {}& - 2\hspace{1pt}\sigma \,  \Delta^{(2)}
+ 2 \hspace{1pt}\sigma \,
\rmd'\delta' + 2 \, \rmd' \delta'\,  \sigma - 4\, \rmd' \sigma\,  \delta' \, ,  \\
\delta_\sigma \, \Delta^{(1)} = {}& - 2\hspace{1pt}\sigma \,  \Delta^{(1)} + 2 \hspace{1pt}\sigma \,
\rmd'\delta' + 4 \, \rmd' \delta'\,  \sigma - 6\, \rmd' \sigma\,  \delta' + 2 \,
\delta' \sigma \, \rmd' - 2\hspace{1pt}\sigma \, \delta' \rmd' \, , \\
\delta_\sigma \, \Delta^{(0)} = {}& - 2\hspace{1pt}\sigma \,  \Delta^{(0)} -4 \hspace{1pt}\sigma \, \delta' \rmd'
+ 4\, \delta' \sigma\,  \rmd' \, .}[]
Using relations such as $\rmd' \Delta^{(1)} = \Delta^{(2)} \rmd'$ we may
obtain
\eqna{
& \delta_\sigma
\Big ( \Tr_{\Omega^{(2)}}\big ( e^{-\tau \, \Delta^{(2)}} \big ) -
2 \, \Tr_{\Omega^{(1)}}\big ( e^{-\tau \, \Delta^{(1)}} \big )
+ 3\, \Tr_{\Omega^{(0)}} \big ( e^{-\tau \, \Delta^{(0)}} \big ) \Big ) \\
&{} = - 2\tau \frac{\rmd}{\rmd \tau} \Big (
\Tr_{\Omega^{(2)}} \big (\sigma\,  e^{-\tau \, \Delta^{(2)}} \big )
- 2 \,  \Tr_{\Omega^{(1)}} \big ( \sigma \, e^{-\tau \, \Delta^{(1)}} \big )
+ 3\, \Tr_{\Omega^{(0)}} \big (\sigma \,  e^{-\tau \, \Delta^{(0)}} \big )
\Big ) \, ,}[]
so that from \eqref{Wtwo} and \eqref{zeta}
\be
\delta_\sigma W =
\Big (  \Tr_{\Omega^{(2)}}\big ( \sigma \, e^{-\tau \, \Delta^{(2)}} \big ) -
2 \, \Tr_{\Omega^{(1)}}\big ( \sigma\, e^{-\tau \, \Delta^{(1)}} \big )
+ 3\, \Tr_{\Omega^{(0)}} \big (\sigma\,  e^{-\tau \, \Delta^{(0)}} \big ) \Big ) \Big |_{\tau^0} \, ,
\label{sW}
\ee
with $|_{\tau^0}$ denoting the $\tau^0$ term in the Laurent expansion in $\tau$.

In each case the Laplacians defined in \eqref{laplace} have the form
\be
\Delta = - \D^2 + 2\, {\hat R}  \, {\mathds{1}}_V  + Y_\Delta \, ,
\label{deltan}
\ee
for $\Delta : V \to V$ and $\D_\mu = \nabla_\mu + A_{\mu}$ with $A_\mu$
an appropriate connection on $V$.
For such elliptic operators the associated
heat kernel $K_\Delta(x,y;\tau)$,  corresponding to $e^{-\tau \Delta}$, has the
well known expansion $(4\pi \tau)^{\frac{1}{2}d} K_\Delta(x,x;\tau) \sim \sum_{n\ge 0}
a_{\Delta,n}|(x) \, \tau^n $ with $a_{\Delta,n}|$ the diagonal Seeley--DeWitt coefficients.
Hence for $d=6$
\be
(4\pi)^3 \, \Tr_V \big ( \sigma \, e^{-\tau \Delta} \big ) \big |_{\tau^0} =
\int \rmd^6x \sqrt{\gamma} \;  \sigma \, \tr_V (  a_{\Delta,3}| ) \, ,
\ee
with $\tr_V$ the matrix trace and
\eqna{7!\; \tr_V(a_{\Delta,3}| )
={}& \dim V \big ( \tfrac{5}{9} \, E_6 - \tfrac{28}{3} \, I_1 + \tfrac{5}{3} \, I_2 + 2\, I_3 \big ) \\
&{}+14\big ( 3\,  \tr_V ( {\hat I} )
+ 5\,  W_{\mu\nu\lambda\rho}\, \tr_V( F^{\mu\nu}F^{\lambda\rho} ) -8 \,
\tr_V(F^{\mu\nu}F_{\nu\lambda}F^\lambda{}_\mu)  \big ) \\
&{} - 7! \, \tr_V \big ( \tfrac{1}{6} \, Y_\Delta{\!}^3 + \tfrac{1}{12} \, Y_\Delta \,
\Delta_2 Y_\Delta  + \tfrac{1}{180} \,
W^{\rho\mu\nu\lambda}  W_{\rho\mu\nu\lambda}\, Y_\Delta
+ \tfrac{1}{12 }\, F^{\mu\nu}F_{\mu\nu} \,Y_\Delta \big ) \\
&{} + 7! \, \nabla_\mu \nabla_\nu Z_\Delta{\vphantom{Z}\!}^{\mu\nu} \, .
}[athree]
Here $ \tr_V ( {\hat I} ) = \tr_V( F^{\mu\nu}\D^2 F_{\mu\nu}) + \cdots,$
with ${\hat I}$ given in \eqref{Finv}, is a dimension six conformal scalar
formed from $F_{\mu\nu}$, and $\Delta_2 = -\D^2 + 2\, {\hat R}$ with
$\D_\mu Y_\Delta = \pr_\mu Y_\Delta + [ A_\mu, Y_\Delta]$. For zero-forms
$F_{\lambda\rho}\to 0$ while acting on one-forms $A_\mu$,
$F_{\lambda\rho}\to R_\mu{}^{\mu'}{\!}_{\lambda\rho}$ and on two-forms
$B_{\mu\nu}$, $F_{\lambda\rho}\to 2\,
\delta_{[\mu}{}^{[\mu'}R_{\nu]}{}^{\nu']}{}_{\lambda\rho}$.  An explicit
form for $Z_\Delta{\vphantom{Z}\!}^{\mu\nu}$ in \eqref{athree} is given in
appendix \ref{conformal}.

For the operators $\Delta^{(n)}$, and letting $Y_{\Delta^{(n)}} \equiv Y_n$,
\eqna{Y_0 = {}& - 2\, {\hat R} + U \, , \\
Y_{1\, \mu}{\vphantom{Y}\!}^{\mu'} = {}&
(- {\hat R} + U)\, \delta_\mu{\!}^{\mu'}
+ 4\, P_\mu{\!}^{\mu'} + U_\mu{\!}^{\mu'} \, , \\
Y_{2\, \mu\nu}{\vphantom{Y}\!}^{\mu'\nu'} = {}&
 U \, \delta_\mu{\!}^{[\mu'} \delta_\nu{\!}^{\nu']}
+ 2\big (2\, P_{[\mu}{\!}^{[\mu'} + U_{[\mu}{\!}^{[\mu'}\big )\,
\delta_{\nu]}{}^{\vphantom{\mu'}\nu']} - W_{\mu\nu}{}^{\mu'\nu'}\,
,}[]
where
\be
U = \frac{1}{2} \, \nabla_\mu v^\mu + \frac{1}{4} \, v_\mu v^\mu \, , \qquad
 U_{\mu\nu} = U_{\nu\mu}=  - \nabla_\mu v_\nu \, , \qquad v_\mu = g^2 \pr_\mu \frac{1}
 {g^2} \, .
 \label{defUv}
\ee
From \eqref{athree} $\delta_\sigma W$ in \eqref{sW} is then determined
in the form \eqref{basic} with
\be
L_6 = \tr_{\Omega^{(2)}} \big ( a_{\Delta^{(2)},3}| \big ) - 2\,
\tr_{\Omega^{(1)}} \big ( a_{\Delta^{(1)},3}| \big ) + 3\,
\tr_{\Omega^{(0)}} \big ( a_{\Delta^{(0)},3}| \big ) +\nabla_\mu\nabla_\nu Z^{\mu\nu}  \, ,
\label{L6two}
\ee
up to the arbitrariness in \eqref{freedom}. Here $\tr_{\Omega^{(2)}}(1)=15, \,
\tr_{\Omega^{(1)}}(1)=6, \, \tr_{\Omega^{(0)}}(1)=1$. The various traces necessary
to determine \eqref{L6two} using \eqref{athree} are given in appendix
\ref{resTwoForms}.
Neglecting the terms involving $U$ we get
\be
L_6^R = \tfrac{1}{7!} \big ( - \tfrac{1}{3} \, 8008 \, I_1 - \tfrac13\, 2378 \, I_2
+ 180\, I_3 + 442 \, E_6 \big ) \, ,
\label{tsey}
\ee
which reproduces the results of \cite{Bastianelli} listed in \eqref{table}.
In terms of $v_\mu$ defined in \eqref{defUv}
\eqna{
L_6^g = {}& -  \tfrac{1}{8} \, \pr^\lambda \nabla_\mu v^\mu \, \pr_\lambda
\nabla_\nu v^\nu + 2 \, P_{\mu\nu}\,  \nabla^\mu v^\nu \,  \nabla_\lambda v^\lambda
- \tfrac{3}{4} \, {\hat R}  \, \nabla_\mu v^\mu\,  \nabla_\nu v^\nu \\
& {} - \big (  B_{\mu\nu}  + 6 \, P_{\mu\lambda}P^\lambda{}_\nu - 4 \,  P_{\mu\nu} {\hat R}
+ \nabla_\mu \pr_\nu {\hat R} \big ) v^\mu v^\nu
+ \big ( P^{\lambda\rho}P_{\lambda\rho} - {\hat R}^2 + \half \, \nabla^2 {\hat R} ) v_\mu v^\mu \\
&{}+ \tfrac13 \, W^{\mu\lambda\rho\omega} W^\nu{}_{\lambda\rho\omega} \, v_\mu v_\nu
- \tfrac{11}{120}\,  W^{\mu\nu\lambda \rho} W_{\mu\nu\lambda\rho} \, v^\omega v_\omega \\
&{} - \quar \big ( ( \nabla^\mu v^\nu\,  \nabla_\mu v_\nu - \quar \, \nabla_\mu v^\mu
\, \nabla_\nu v^\nu )\,  v^\rho v_\rho + {\hat R} \, (v^\rho v_\rho)^2 \big ) \\
&{}
- \tfrac{1}{32} \, v^\mu v_\mu \, \Delta_2 (v^\nu v_\nu)
- \tfrac{1}{64} \, (v^\mu v_\mu)^3 \, ,}[L6g2]
for $\Delta_2 = - \nabla^2 + 2\, {\hat R}$. The first two lines in \eqref{L6g2}
agree with the form expected from $S_1 + S_2 +S_3$ given by
\eqref{S1}, \eqref{S2} and \eqref{S3} and there are also contributions which
may be identified with $j_2, j_3$ in \eqref{T1}, \eqref{T2}, with coefficients $-\tfrac{1}{16},
\, - \quar$,  as well as \eqref{X6k}.

\newsec{Calculations in Scalar \texorpdfstring{$\phi^3$}{phi3}
Theory}[phicubed]

In six dimensions the only conventionally renormalisable quantum field theory
is the apparently unphysical  (although for imaginary couplings
the theory has relevance in statistical physics \cite{Fisher}) $\phi^3$ theory
given by the Lagrangian
\be
{\mathscr L} (\phi,V) = - \half \big ( \pr^\mu \phi_i \pr_\mu \phi_i +
\xi_d \, \phi_i \phi_i \, R \big ) - V(\phi) \, , \quad V(\phi)=
 \tfrac{1}{6} \, \gl_{ijk} \, \phi_i \phi_j \phi_k \, , \quad i=1,\dots, n_\phi \, ,
\label{phi3}
\ee
where Weyl invariance in six dimensions requires $\xi_6 = \tfrac{1}{5}$. However
using dimensional regularisation with $d=6-\vep$ it is necessary
to keep $\vep$-dependent terms to ensure compatibility with conformal constraints
to two loop order so that $\xi_d = \tfrac{1}{5} - \tfrac{1}{100} \vep + \rO(\vep^2)$.
Two loop calculations for six dimensional $\phi^3$ theory on curved backgrounds were
initiated in \cite{Jack, Kodaira} and recently extended to local couplings in
\cite{asix} while the $\beta$-function has been determined to three loops in
\cite{Mckane1, Mckane2, Kleb}.

For a finite perturbative expansion  starting from \eqref{phi3} it is necessary
of course to add counterterms  ${\mathscr L}_{\rm c.t.}$ containing poles in $\vep$.
These may be restricted  to the form, up to total derivatives,
\be
{\mathscr L}_{\rm c.t.}(\phi,V) \equiv
- \half\,  \tr \big ( \pr^\mu \tphi \, N \pr_\mu \tphi+ \xi_d \,
\tphi \, N  \tphi \, R \big )
- V_{\rm c.t.}(\tphi) \, , \qquad \tphi_{ij} = \gl_{ijk} \phi_k\, .
\label{CT}
\ee
$V_{\rm c.t.}(\tphi)$ is a polynomial of degree three and
 includes $\phi$-independent terms of dimension six
depending on the curvature and derivatives of the couplings. Renormalisability
on a curved background and with local couplings
dictates that in \eqref{phi3} ${\mathscr L}(\phi,V)$ should be extended to
${\mathscr L}(\phi,V,a)$ depending on
a background gauge field $a_{\mu\, ij} = - a_{\mu\, ji} $ and also a general cubic $V$,
\be
\pr_\mu \phi_i \to (D_\mu \phi)_i = \pr_\mu \phi_i + a_{\mu\, ij} \phi_j \, , \quad
V(\phi)  = \tfrac{1}{6} \, \gl_{ijk} \, \phi_i \phi_j \phi_k +\half \,
m_{ij} \, \phi_i \phi_j + h_i \, \phi_i \, ,
\ee
so that
\be
{\mathscr L}_0 \equiv  {\mathscr L}(\phi,V,a) + {\mathscr L}_{\rm c.t.}(\phi,V,a) =
{\mathscr L}(\phi_0,V_0,a_0) - \frac{1}{(4\pi)^3} \, \chi(V,a) \, .
\label{Lzero}
\ee
Here $\chi$ is a  dimension six scalar independent of $\phi$ and formed from
the curvature and the couplings with  derivatives.

The RG equations take the form
\be
(4\pi)^{\frac{1}{2}d} \big ( \delta_\sigma + d \, \sigma + \D_\beta + \D_\phi \big ) {\mathscr L}_0 =
\sigma \, L_6 + \nabla_\mu  ( \X^{\mu\nu} \pr_\nu \sigma ) \, ,
\label{RGeq}
\ee
for
\eqna{\D_\beta = {}&  \int \! \rmd^d x \; \sigma \bigg (
{\hat \beta}_{\lambda\, ijk} \,  \frac{\delta}{\delta \lambda_{ijk}}
+ {\hat \beta}_{m\, ij}  \, \frac{\delta}{\delta m_{ij}}
+ {\hat \beta}_{h\, i} \,  \frac{\delta}{\delta h_{i}}
+ (\rho \cdot D_\mu \lambda)_{ij}  \, \frac{\delta}{\delta a_{\mu\,  ij}}
\bigg ) \, , \\
\D_\phi = {}&  - \int \! \rmd^d x \; \sigma \big ( \half (d-2) \delta_{ij} + \gamma_{ij} \big )
\phi_j \, \frac{\delta}{\delta \phi_{i}} \, .}[Dbeta]
In \eqref{Dbeta} ${\hat \beta}_{\lambda\, ijk} = - \half \vep \, \lambda_{ijk}
+ {\beta}_{\lambda\, ijk} , \,  {\hat \beta}_{m\, ij}  = - 2\, m_{ij} + (\gamma_m \cdot m)_{ij}
+ {\beta}_{m\, ij} $, with ${\beta}_{m\, ij}$ independent of $m$,
and ${\hat \beta}_{h\, i}  = -  \big ( \half (d + 2) \delta_{ij} - \gamma_{ij} \big )
h_j +  \beta_{h\, i}$, with $\beta_h$ independent of $h$.
 $\D_\beta$ may contain additional terms involving $\pr_\mu \sigma$
but these are neglected as they are unimportant here. As usual \eqref{RGeq} determines
the higher order $\vep$ poles in ${\mathscr L}_0$.

As shown by Brown and Collins in four dimensions for $\phi^4$
theory \cite{Brown} the subtraction prescription implied by \eqref{CT}
suffices to ensure Weyl invariance remains valid to  one loop order so that
results at two loops for the $\phi$-independent counterterms should be consistent
with the general constraints described here.
At one loop the necessary  counterterms are determined by $a_{\Delta,3}|$ for the
operator $\Delta = (- \nabla^2  + \half (d-2) {\hat R}) 1 + m +  \tphi$ which gives
\eqna{(4\pi)^3  V_{\rm c.t.}(\tphi)^{(1)} ={}&  \frac{1}{\vep} \Big (
- \tfrac{1}{6} \, \tr \big ( (m+\tphi)^3\big )
 -  \tfrac{1}{180} \,  \tr (m+ \tphi)\,  W^{\rho\mu\nu\lambda}  W_{\rho\mu\nu\lambda} \\
 \noalign{\vskip - 2pt}
 &{} \hskip 1cm + \tfrac16 \, \tr \big ( \tphi \, (\nabla^2 - 2\, {\hat R} )m  \big )
 -  \tfrac{1}{12} \, \tr \big ( \pr^\mu m \, \pr_\mu m + 2\, {\hat R} \, m^2 \big ) \\
\noalign{\vskip - 2pt}
&{}\hskip 1cm+ n_\phi\,  \tfrac{1}{7!}\big (
\tfrac{5}{9} \, E_6 - \tfrac{28}{3} \, I_1 + \tfrac{5}{3} \, I_2 + 2\, I_3 \big ) \Big ) \, , \\
( 4\pi)^3 N^{(1)} = {}& - \frac{1}{6\vep} \, .}[oneloop]
The one loop results for $\beta_\lambda, \gamma$ are standard, and are given in \cite{asix},
but in addition we must take $(\rho^{(1)} \cdot \rmd \lambda)_{ij} = - \tfrac{1}{12}( \lambda_{ikl} \,
\rmd \lambda_{jkl} - \rmd \lambda_{ikl} \, \lambda_{jkl} )$,
$\beta^{(1)}{\!}_{m\, ij} = -  \tfrac16 \, \pr^\mu \lambda_{ikl} \, \pr_\mu \lambda_{jkl}$
as well as $\beta^{(1)}{\!}_{h\, i} |_{m=0} =  - \tfrac{1}{180} \, \lambda_{ijj} \,
W^{\rho\mu\nu\lambda}  W_{\rho\mu\nu\lambda}$.
For the scalar theory defined by \eqref{phi3} it is  then easy to read off
\be
L_6^{R\, (1)} =   n_\phi\,  \tfrac{1}{7!}\big (
\tfrac{5}{9} \, E_6 - \tfrac{28}{3} \, I_1 + \tfrac{5}{3} \, I_2 + 2\, I_3   \big ) \, ,
\label{1loop}
 \ee
which of course confirms the results for free scalar fields in \eqref{table}.

Extending the calculations  to two loops, letting $\gl_{ijk} \to
(4\pi)^{\frac{3}{2}} \gl_{ijk}$, leads to
\be
L_6^{R\, (2)} =  \frac{\gl_{ijk}\gl_{ijk}}{9\times6!}\, \Big (
 \tfrac{2}{9} \, I_1 - \tfrac{13}{18} \, I_2 - \tfrac{1}{4}\, I_3   \Big ) \, .
\label{2loop}
\ee
This is in agreement with similar two loop calculations in
\cite{Jack,Kodaira}\footnote{In \cite{Jack} the relevant results are
contained in (3.21) but it is necessary to have an additional factor $\vep$ in
the $R^2(\xi R + \dots )$ term.}
although a non conformal tensorial basis was used in these papers.

The  two loop calculations may also be extended to allow for $x$-dependent couplings
leading to contributions to $\chi^{(2)}$ in \eqref{Lzero} involving derivatives of $\gl$.
There is a single double pole in $\vep$, independent of $\phi$, which is
 proportional to $ \pr^\omega \lambda_{ijk}
\pr_\omega \lambda_{ijk} \, W^{\rho\mu\nu\lambda}  W_{\rho\mu\nu\lambda}$ whose
coefficient is in accord with \eqref{RGeq}, although it is necessary to take account of the $m$
terms in \eqref{oneloop}.
Discarding terms with two overall derivatives and also some scheme dependent terms
proportional to $ W^{\rho\mu\nu\lambda}  W_{\rho\mu\nu\lambda}$
these may be reduced to a conformally covariant form and give, after rescaling $\gl$ as
before to absorb factors of $4\pi$,
\eqna{L_6^{g\, (2)} = {}& - \tfrac{1}{6\times 6!} \big ( \pr^\mu \nabla^2 \gl_{ijk} \,
\pr_\mu \nabla^2 \gl_{ijk}
- 16 \, P^{\mu\nu} \,  \nabla^2 \gl_{ijk} \, \nabla_\mu \pr_\nu  \gl_{ijk}
+ 6 {\hat R} \, \nabla^2 \gl_{ijk} \,\nabla^2 \gl_{ijk} \\
& \hskip 1.3cm {}+ 8 (  B^{\mu\nu}
+ 6\, P^{\mu\lambda}P^\nu{\!}_\lambda - 4\, P^{\mu\nu}{\hat R}
+ \nabla^\mu \pr^\nu {\hat R} ) \, \pr_\mu  \gl_{ijk} \, \pr_\nu  \gl_{ijk} \\
& \hskip 1.3cm {} -4  (  2\, P^{\lambda\rho}
P_{\lambda\rho} - 2 \, {\hat R}^2  +  \, \nabla^2 {\hat R}  ) \,
\pr^\mu  \gl_{ijk} \, \pr_\mu  \gl_{ijk} \big ) \\
&{}- \tfrac{4}{9\times 6!} \, W^{\mu\lambda\rho\omega} W^\nu{}_{\lambda\rho\omega}
\,  \pr_\mu  \gl_{ijk} \, \pr_\nu  \gl_{ijk}  
- \tfrac{23}{180\times 6!} \, W^{\nu\lambda\rho\omega} W_{\nu\lambda\rho\omega}
\,  \pr^\mu  \gl_{ijk} \, \pr_\mu  \gl_{ijk} \, .}[L6gphi]
This has exactly the form expected from  \eqref{L6g} and shows the presence of
all three possible two index tensors on the conformal manifold although the coefficient of
the last term in \eqref{L6gphi} is scheme dependent.

\newsec{Positivity Constraints}[positivity]

The various terms present in $L_4,L_6$ correspond to contact terms for
identities resulting from Weyl scaling for correlation functions of the
operators $\O_I$ coupled to the marginal couplings $g^I$ and also the
energy momentum tensor. Positivity conditions arise most straightforwardly
by considering two point functions. Restricting $\sigma$ to be a constant
then \eqref{basic} is equivalent to
\be
(4\pi)^{\frac{d}{2}} \, \mu \frac{\pr}{\pr\mu} W = \int \rmd^d x \sqrt{\gamma}\; L_d \, ,
\label{basic2}
\ee
for $\mu$ a regularisation scale and where, by analytic continuation, the metric
is taken to be Euclidean and $iW\to W$. Applied to the two point function, obtained
by functional differentiation of $W$ twice with respect to $g$, \eqref{basic2} requires
\be
\mu \frac{\pr}{\pr\mu} \big \langle \O_I(x) \, \O_J(0) \big \rangle
\big |_{\pr g=0, \gamma_{\mu\nu} = \delta_{\mu\nu}} =
\begin{cases} \displaystyle g_{IJ}\, ( \pr^2)^2 \delta^4(x) / (4\pi)^2 \, ,  &  d=4  \, ,\\
\displaystyle  g_{IJ} \,  ( \pr^2)^3 \delta^6(x) / (4\pi)^3 \, ,  &  d=6  \, .
\label{Ward}
\end{cases}
\ee
Conformal invariance dictates
\be
 \big \langle \O_I(x) \, \O_J(0) \big \rangle
\big |_{\pr g=0, \gamma_{\mu\nu} = \delta_{\mu\nu}} =
 G_{IJ}\; \R \frac{1}{(x^2)^d} \, .
 \label{Rdef}
\ee
For general $d$, $(x^2)^{-\alpha}$ may be defined as an  analytic function  in
$\alpha$ with poles at $\alpha = \half  d + n, \ n=0,1,2,\dots$. Hence for
$d$ even it is necessary to regularise, denoted in \eqref{Rdef} by $\R$,
 so that $(x^2)^{-d}$ makes sense as
a distribution for all $x$, or equivalently has a well defined Fourier
transform. This is essential in order to make a connection with the identities
in  \eqref{Ward} and requires the introduction of the arbitrary scale $\mu$. A
convenient prescription is provided by differential regularisation
\cite{FreedmanD}, which gives
\eqna{\R \frac{1}{(x^2)^4}  ={}&  - \frac{1}{4^4 \times 3} \, (\pr^2)^3
\Big ( \frac{1}{x^2} \, \ln \mu^2 x^2 \Big ) \, , \quad d=4 \, , \\
\R \frac{1}{(x^2)^6}  =  {}&
- \frac{1}{4^6 \times 45} \, (\pr^2)^4 \Big ( \frac{1}{(x^2)^2} \, \ln \mu^2 x^2 \Big )
\, , \quad d=6 \,  .}[dist]
Substituting \eqref{Rdef} with \eqref{dist}  on the left hand side of \eqref{Ward} gives
\be
(2\pi^2)^2 G_{IJ} = 24\, g_{IJ} \, , \quad d=4\, , \qquad
\pi^6 G_{IJ} = 360\, g_{IJ} \, , \quad d=6 \, .
\label{Ggrel}
\ee

Unitarity implies positivity conditions on $G_{IJ}$. To apply unitarity
here it is sufficient to use the Fourier transforms
\eqna{
\int \rmd^4 x \; e^{i k \cdot x}\frac{1}{x^2} \, \ln \mu^2 x^2 = {}& -
\frac{4\pi^2}{k^2} \, \ln \frac{e^{2\gamma} k^2}{4\mu^2} \, , \\
\int \rmd^6 x \; e^{i k \cdot x}\frac{1}{(x^2)^2} \, \ln \mu^2 x^2 = {}& -
\frac{4\pi^3}{k^2} \,\Big ( \ln \frac{e^{2\gamma} k^2}{4\mu^2} - 1 \Big )
\, ,
}[FTx]
where $\gamma$ is the Euler--Mascheroni constant. It is then
straightforward from \eqref{FTx} to determine the Fourier transforms of $\R
\frac{1}{(x^2)^4}$, $\R \frac{1}{(x^2)^6}$ as given by \eqref{dist} for
$d=4,6$. Under analytic continuation from Euclidean to Minkowski space $k_d
\to -i k_0$ and the absorptive part for $k^2<0$ is given by
$\operatorname{Im} \ln (k^2 - i \epsilon) = - \pi \, \theta(-k^2)$.
Applied to \eqref{Rdef} this requires positivity of $G_{IJ}$.

For free scalar theories $(4\pi)^3 \langle \tfrac16 \phi^3(x) \, \tfrac16 \phi^3(0) \rangle
= 1/(6\pi^6 (x^2)^6)$ so that in \eqref{Rdef} we may take $G_{IJ} = \delta_{IJ}/(6\pi^6)$.
Using \eqref{Ggrel} $g_{IJ}= \delta_{IJ}/(3\times 6!)$ in agreement with
\eqref{L6gphi}. For two-forms, from \eqref{twoL}, \eqref{twoGF},
\eqna{\big \langle B^{\mu\nu}(x) \, B_{\lambda\rho}(0) \big \rangle ={}&
\frac{g^2}{2\pi^3} \, \delta^{[\mu}{\!}_\lambda \delta^{\nu]}{\!}_\rho \, \frac{1}{(x^2)^2} \, , \\
\big \langle (\rmd B)^{\mu\nu\omega}(x) \, {\rmd B}_{\lambda\rho\sigma}(0) \big \rangle ={}&
\frac{18 \,g^2}{\pi^3} \, I^{[\mu}{\!}_\lambda(x) I^{\nu}{\!}_\rho(x) I^{\omega]}{\!}_\sigma (x) \,
\frac{1}{(x^2)^3} \, ,}[]
where
\be
 I_{\mu\nu}(x) = \delta_{\mu\nu} - 2 \, \frac{x_\mu x_\nu}{x^2}
\ee
is the inversion tensor. In this case for $\O = \tfrac{1}{12}
(\rmd B)^{\mu\nu\omega} (\rmd B)_{\mu\nu\omega}$ then $\pi^6 G_{IJ} \to 90\, g^4$
so that $g_{IJ} \to \quar \, g^4$. This is in agreement with \eqref{L6g2}.

These considerations may also be applied to the energy momentum tensor
defined by functional differentiation with respect to the metric. For the
two point function only the Weyl anomaly proportional to $c$ in \eqref{Lfour}
contributes to the corresponding equation to \eqref{Ward} when $d=4$; for
$d=6$ just the term
$W^{\rho\mu\nu\lambda} \nabla^2  W_{\rho\mu\nu\lambda}$, contained in  $I_3$
and  proportional to $c_3$, in \eqref{Lsix} is relevant. Thus
\be
\mu \frac{\pr}{\pr\mu} \big \langle T_{\mu\nu}(x) \, T_{\sigma\rho}(0) \big \rangle
\big |_{\pr g=0, \gamma_{\mu\nu} = \delta_{\mu\nu}} =
\begin{cases} \displaystyle 4 c\; \D_{\mu\nu\sigma\rho}\,  \delta^4(x) / (4\pi)^2 \, ,
&  d=4  \, ,\\
\displaystyle  6 c_3 \;  \D_{\mu\nu\sigma\rho} \, \pr^2 \delta^6(x) / (4\pi)^3 \, ,  &
d=6  \, ,
\label{WardT}
\end{cases}
\ee
where, for general $d$,
\be
\D_{\mu\nu\si\rho} = \half \bigl ( S_{\mu\si} S_{\nu\rho} +
S_{\mu\rho} S_{\nu \si} \bigl ) {} - \frac{1}{d-1}\, S_{\mu\nu}S_{\si\rho}\, , \quad
S_{\mu\nu} =  \pr_\mu\pr_\nu - \delta_{\mu\nu} \pr^2 \, .
\ee
For conformal theories
\be
 \big \langle T_{\mu\nu}(x) \, T_{\sigma\rho}(0)  \big \rangle
\big |_{\pr g=0, \gamma_{\mu\nu} = \delta_{\mu\nu}} =
 C_T \; \R \bigg (  \frac{1}{(x^2)^d} \, \I_{\mu\nu\sigma\rho}(x) \bigg ) \, ,
  \label{Tdef}
 \ee
 with the inversion tensor for symmetric traceless rank two tensors
 \be
 \I_{\mu\nu\sigma\rho} =  \half \bigl ( I_{\mu\si}I_{\nu\rho} +
I_{\mu\rho} I_{\nu \si} \bigl ) {} - \frac{1}{d}\, \delta_{\mu\nu} \delta_{\si\rho} \, .
\ee
Since
\be
 \D_{\mu\nu\si\rho} \frac{1}{(x^2)^{d-2}} = 4(d-2)^2 d(d+1)\,
\frac{1}{(x^2)^d} \, \I_{\mu\nu\sigma\rho}(x)  \, ,
\ee
then in \eqref{Tdef} we may define
\eqna{\R \bigg (  \frac{1}{(x^2)^4} \, \I_{\mu\nu\sigma\rho}(x) \bigg )
= {}& - \frac{1}{4^4 \times 5} \,
 \D_{\mu\nu\sigma\rho} \, \pr^2 \Big ( \frac{1}{x^2} \, \ln \mu^2 x^2 \Big )  \, ,
\quad d=4 \, , \\
 \R \bigg (  \frac{1}{(x^2)^6} \, \I_{\mu\nu\sigma\rho}(x) \bigg )
= {}& - \frac{1}{4^6 \times 63} \,  \D_{\mu\nu\sigma\rho} \, (\pr^2)^2
\Big ( \frac{1}{(x^2)^2} \, \ln \mu^2 x^2 \Big )  \, , \quad d=6 \,.}[FFT]
Hence
\be
(2\pi^2)^2 C_T  = 160 \, c  \, , \quad d=4\, , \qquad
\pi^6   C_T = \tfrac{3}{5} \times 7! \, c_3  \, , \quad d=6 \, .
\ee
The relation between $C_T$ and $c$ for $d=4$ was obtained in \cite{Pet}
and the connection between $C_T$ and $c_3$ for $d=6$ in \cite{Bastianelli}.
For $d=6$ the results in \eqref{table} are in agreement with calculations
of $C_T$ for scalars, fermions in \cite{Pet} and also two-form gauge fields in
\cite{Buchel}. The results \eqref{FTx} suffice to determine the Fourier
transforms of \eqref{FFT}. Under continuation to Minkowski space we
must take $T_{di} \to -i T_{0i}, \, i=1,\dots, d-1, \, T_{dd} \to - T_{00}$, so that
in \eqref{Tdef}
$\delta_{\mu\nu} \to \eta_{\mu\nu}$. It follows directly that unitarity
requires $C_T>0$.

Positivity conditions for conserved vector currents $V_{a\mu}$ may be
obtained in a similar fashion. Correlation functions containing $V_{a\mu}$
are defined by functional differentiation of $W$ with respect to a
background gauge field $A_{a\mu}$. Then, from
\eqref{Lfour} and \eqref{LsixF}, taking
$L_6^F \to - \quar \, \kappa_{ab} \,  F_a{\!}^{\mu\nu}\,
\nabla^2 F_{b\,\mu\nu}$,
\be
\mu \frac{\pr}{\pr\mu} \big \langle V_{a \mu}(x) \, V_{b\nu}(0) \big \rangle
\big |_{\pr g=0, \gamma_{\mu\nu} = \delta_{\mu\nu}} =
\begin{cases} \displaystyle
- \kappa_{ab} \; S_{\mu\nu}\,  \delta^4(x) / (4\pi)^2 \, ,  &  d=4  \, ,\\
\displaystyle
-
\kappa_{ab} \;  S_{\mu\nu} \, \pr^2 \delta^6(x) / (4\pi)^3 \, ,  &  d=6  \, .
\label{WardV}
\end{cases}
\ee
For conformal theories the vector two point function has the form
\be
 \big \langle V_{a\mu}(x) \, V_{b\nu}(0)  \big \rangle
\big |_{\pr g=0, \gamma_{\mu\nu} = \delta_{\mu\nu}} =
C_{V ab} \; \R \bigg (  \frac{1}{(x^2)^{d-1} }\, I_{\mu\nu}(x) \bigg ) \, .
\label{Vdef}
\ee
 In this case
 \be
 S_{\mu\nu} \frac{1}{(x^2)^{d-2}} = -2(d-2)(d-1)\,
  \frac{1}{(x^2)^{d-1}} \, I_{\mu\nu}(x)  \, ,
 \ee
so that in \eqref{Vdef} we may take
\eqna{\R  \bigg (  \frac{1}{(x^2)^3} \, I_{\mu\nu}(x) \bigg )
= {}&  \frac{1}{48} \,
 S_{\mu\nu} \, \pr^2 \Big ( \frac{1}{x^2} \, \ln \mu^2 x^2 \Big )
\, , \quad d=4 \, ,  \\
 \R  \bigg (   \frac{1}{(x^2)^5} \,  I_{\mu\nu}(x) \bigg )
 = {}& \frac{1}{3840} \,
 S_{\mu\nu} \, (\pr^2)^2 \Big ( \frac{1}{(x^2)^2} \, \ln \mu^2 x^2 \Big )
 \, , \quad d=6 \, .}[]
Hence \eqref{WardV} requires
\be
(2\pi^2)^2 C_{Vab}  = \tfrac{3}{2} \, \kappa_{ab}  \, , \quad d=4\, , \qquad
\pi^6   C_{Vab} = \tfrac{15}{2}  \, \kappa_{ab}  \, , \quad d=6 \, .
\ee
The results for $\kappa$ in \eqref{table} agree with $C_V$ calculated for
free scalars and fermions in \cite{Pet}.

There are further positivity constraints on the energy momentum tensor
three point function which arise by requiring that the energy flux in light-like
directions  must be positive \cite{Hofman}. For $d=6$ the conditions
take the form \cite{deBoer}
\eqna{C_1 \equiv {}& 1 - \tfrac{1}{5} t_2 - \tfrac{2}{35} t_4 \ge 0 \, , \quad
C_2 \equiv 1 - \tfrac{1}{5} t_2 - \tfrac{2}{35} t_4 + \half t_2 \ge 0 \, , \\
C_3 \equiv {}&  1 - \tfrac{1}{5} t_2 - \tfrac{2}{35} t_4 + \tfrac{4}{5}(t_2+t_4) \ge 0 \, ,}[ineq]
with $t_2,t_4$ corresponding to the possible angular dependencies
of the energy flux at null infinity. $t_2,t_4$ depend on the three possible
structures for the conformal energy momentum tensor three point function
after factoring $C_T$ as determining the overall normalisation.
In six dimensions these are determined by the coefficients $c_1,c_2,c_3$
in the conformal anomaly \eqref{Lsix} (unlike in four dimensions $a$ is
irrelevant as far as the energy momentum tensor three point function is
concerned). It is sufficient to use the results for free fields in \cite{deBoer} and
\eqref{table} which give in general
\be
t_2 = \frac{15(23\, c_1-44\, c_2+ 144 \, c_3)}{16\, c_3} \, , \qquad
t_4 = - \frac{105( c_1 - 2\,  c_2 + 6\,  c_3)}{2\,  c_3} \, .
\label{t24}
\ee
Then from \eqref{ineq} we may obtain, since $c_3>0$,
\eqna{&-21\, c_1 + 36\, c_2 -128 \, c_3 \ge 0 \, , \quad
101\, c_1 -196 \, c_2 +\tfrac{1904}{3} \, c_3 \ge 0 \, , \\
& - 139 \, c_1 + 284 \, c_2 - \tfrac{2432}{3} \, c_ 3 \ge 0 \, .}[]
The inequalities \eqref{ineq} define a triangular region in which the
three free theory results correspond to the vertices where in each
case two different inequalities become equalities.

For free scalars $C_1=C_2 = 0$. It is then non trivial that any conformal
perturbation of a scalar theory should satisfy  the inequalities \eqref{ineq}.
If we use the results for $c_1,c_2,c_3$ provided by \eqref{1loop} and
\eqref{2loop} for $\phi^3$ theory with \eqref{t24} we get
\be
C_1 = \frac{7}{216}\, \gl_{ijk}\gl_{ijk} \, , \qquad
C_2 = \frac{7}{36}\, \gl_{ijk}\gl_{ijk} \, ,
\ee
so that the perturbative corrections respect the inequalities even though this
theory remains potentially sick.

\newsec{Discussion}

The calculations in this paper show that there are significant differences between six
and four dimensions and also two for which Zamolodchikov first derived the $c$-theorem.
In  two dimensions the result for the response to a Weyl rescaling in \eqref{basic}
becomes simply
\be
L_2 = \tfrac{1}{6} \, c \, R -  \half \, g_{IJ} \, \pr^\mu g^I \pr_\mu g^J \, .
\ee
In this case the consistency conditions away from a conformal fixed point essentially
imply
\be
\tfrac{1}{3} \, \pr_I c = g_{IJ} \beta^J \, ,
\ee
which implies irreversibility of RG flow, a strong version of the $c$-theorem, if $g_{IJ}$
is positive definite. In this case positivity holds since $g_{IJ}$ can be related directly
to the two-point function for the operators $\O_I$ coupled  to $g^I$. In four dimensions
away from a fixed point there is no longer a single rank two tensor; in \eqref{Lfour}
the corresponding contributions become
$\half \, a_{IJ} \, \nabla^2 g^I  \nabla^2 g^J - G_4\vphantom{G}^{\mu\nu}\, g_{IJ} \, \pr_\mu  g^I \pr_\nu g^J
-  {\hat R} \, f_{IJ}\,  \pr^\mu g^I  \pr_\mu g^J $. In this case
consistency conditions require
\be
\tfrac{1}{4} \, \pr_I a = g_{IJ} \beta^J \, .
\ee
Only in the neighourhood of a conformal fixed point, when $a_{IJ} = f_{IJ} = g_{IJ}$, does
positivity of the two-point function, linked to $a_{IJ}$, imply positivity of $g_{IJ}$.

In six dimensions the results obtained in \eqref{L6g} show already that even at a conformal
fixed point there are three two-index tensors. Away  from a fixed point the RG flow equation
becomes
\be
\tfrac{1}{12} \, \pr_I a = g_{1,IJ} \beta^J \, ,
\ee
involving $g_{1,IJ}$, which away from the conformal point corresponds the contribution
involving $G_6\vphantom{G}^{\mu\nu} \sim W^{\mu\lambda\rho\omega} W^\nu{}_{\lambda\rho\omega}$,
rather than $g_{IJ}$ which is related to the positive two point function.
Hence, there
are no straightforward positivity restrictions on $g_{1,IJ}$ even near a fixed point.
As shown by \eqref{L6gphi} $g_{1,IJ}$ is negative for $\phi^3$ theory, which reproduces
the challenge to a six dimensional $a$-theorem observed in \cite{asix}. In
contrast,
the calculations for the two-form case in \eqref{L6g2} give a positive result $g_{1,IJ}$.
However, we should note that there is at present no argument implying that $a>0$ in
six dimensions, unlike that given in \cite{Hofman} for the four dimensional $a$.
In six dimensions $a$ is related to the energy momentum tensor four point function
whose analysis is much harder than the three point function considered in \cite{Hofman}.
Of course with supersymmetry there may be further relations between tensor structures
which might link $g_{1,IJ},g_{2,IJ}$ with $g_{IJ}$.

In this paper we have focussed on solutions of the Weyl consistency conditions of the
form given by \eqref{basic}, \eqref{consis}. Additional contributions to $\delta_\sigma W$
may be obtained by considering variations such that
\be
(4\pi)^{\frac{d}{2}} \, \delta_{\sigma} W = \int \rmd^d x \sqrt{-\gamma}\;
\pr_\mu \sigma \,  Y_d\vphantom{Y}^\mu  \, ,
\label{Wvar}
\ee
where, if $Y_d{\!}^\mu$ is a total derivative, then it can generally
be cancelled by local contributions to $W$. Alternative solutions of the
consistency conditions may be obtained if $Y_d\vphantom{Y}^\mu$ satisfies
\eqna{
\delta_{\sigma} Y_d\vphantom{Y}^\mu + d \, \sigma \, Y_d\vphantom{Y}^\mu
={}&  \Y_d\vphantom{\Y}^{\mu\lambda\rho} \, \nabla_\rho \pr_\lambda \sigma + \nabla_\rho
\big ( \Y_d\vphantom{\Y}^{\mu\lambda\rho} \, \pr_\lambda \sigma \big )
+ \E_d\vphantom{E}^{\mu\lambda}\, \pr_\lambda \sigma \, ,  \\
\Y_d {}^{\mu\lambda\rho} = {}& - \Y_d\vphantom{Y}^{\lambda\mu\rho} \, , \qquad
 \E_d\vphantom{E}^{\mu\lambda} =  \E_d\vphantom{E}^{\lambda\mu}\, .}[consis2]
Of course contributions to $ \E_d\vphantom{E}^{\mu\lambda}$ of the form of
$ X_d\vphantom{X}^{\mu\lambda}$ as in \eqref{consis} may be discarded.
In two and four dimensions  examples are given by
\be
Y_2\vphantom{Y}^\mu = - w_I \, \pr^\mu g^I \, , \qquad
Y_4\vphantom{Y}^\mu = -2\, G_4\vphantom{G}^{\mu\nu} w_I \, \pr_\nu g^I + 2\, \pr_{[ I} w_{J]} \,
\pr^\mu g^I \nabla^2 g^J \, ,
\label{w24}
\ee
where $w_I \rmd g^I $ is a one-form and in the four-dimensional case we make use of
\eqref{G4var}. In this case $ \Y_4 {}^{\mu\lambda\rho} = 2 (\gamma^{\mu\rho}
\gamma^{\lambda\nu} - \gamma^{\mu\nu} \gamma^{\lambda \rho}) w_I \pr_\nu g^I$,
$\E_4\vphantom{E}^{\mu\lambda}=0$.
In  \eqref{w24} the normalisations have been chosen to agree with previous
conventions.

In six dimensions it is sufficient to take
\eqna{
Y_6\vphantom{Y}^\mu = {}&  G_6\vphantom{G}^{\mu\nu} w_I \, \pr_\nu g^I \\
&{} + \pr_{[I} w_{J]} \big (-  \tfrac{10}{3} \, W^{\mu\lambda\rho\nu} \,
\pr_\nu g^I \nabla_\lambda \pr_\rho g^J \\
& \hskip 1.6 cm {} + 6 \, P^{\mu\nu}\, \pr^\rho g^I \nabla_\nu \pr_\rho g^J +
6\, P^{\rho\nu} \, \pr_\rho g^I \nabla^\mu \pr_\nu g^J
- 3 \, P^{\mu\nu} \, \pr_\nu g^I \nabla^2 g^J \\
& \hskip 1.6 cm {}  - 6\,  {\hat R}\, \pr^\nu g^I \nabla^\mu \pr_\nu g^J
- \tfrac32 \, \nabla^\mu \pr^\nu g^I \pr_\nu \nabla^2 g^J
+ \tfrac34\, \nabla^2 g^I \pr^\mu \nabla^2 g^J\,  \big ) \\
&{} + \pr_K  \pr_{[I} w_{J]} \big (  \quar \, \nabla^2 g^K \nabla^2 g^I \pr^\mu g^J
- \nabla^\nu \pr^\rho g^K \nabla_\nu \pr_\rho g^I \pr^\mu g^J
- 2\, \nabla^\mu \pr^\nu g^K \nabla^\rho \pr_\nu g^I  \pr_\rho g^J \big ) \, .
}[Ysixsixd]
This satisfies \eqref{consis2} with
\eqna{
\Y_6\vphantom{\Y}^{\mu\lambda\rho} ={}&  - 2 \,
H_6{\vphantom{H}}^{\mu\lambda\rho\nu} w_I \pr_\nu g^I
+ 6 \,  \pr_{[I} w_{J]} \,  \gamma^{\rho[\mu} \, \nabla^{\lambda]} \pr^\nu g^I \pr_\nu g^J \, , \\
\noalign{\vskip 1pt}
\E_6\vphantom{E}^{\mu\lambda} ={}& - 3  \,  \pr_{[I} w_{J]} \big ( \half \,  \gamma^{\mu\lambda} \,
\pr^\nu \nabla^2 g^I \pr_\nu g^J + \pr^{(\mu} \nabla^2 g^I \pr^{\lambda)} g^J \big ) \\
\noalign{\vskip -1pt}
&{}+ 2\, \pr_K  \pr_{[I} w_{J]} \big (  \gamma^{\mu\lambda}\, \pr^\nu g^K \nabla^\rho \pr_\nu
 g^I \pr_\rho g^J + \pr^\rho g^K \nabla^{(\mu} \pr_\rho g^I \pr^{\lambda)} g^J
 +  3\, \nabla^{(\mu} \pr^\rho g^K \pr_\rho g^I \pr^{\lambda)} g^J \big ) \, ,
 }[calYEsixd]
where $H_6\vphantom{H}^{\mu\lambda\rho\nu}$ is defined by \eqref{Hsix} for
$d=6$. We note that
\eqna{Y_6^\prime\vphantom{Y}^\mu= w_{KIJ}\big(
&\tfrac14 \, \nabla^2 g^K\nabla^2 g^I\pr^\mu g^J- \nabla^\nu\pr^\rho g^K
\nabla_\nu\pr_\rho g^I\pr^\mu g^J\\
{}& +\nabla^\mu\pr^\nu g^K\nabla^\rho\pr_\nu g^I\pr_\rho g^J
-2\,\nabla^\rho\pr^\nu g^K\nabla^\mu\pr_\nu g^I\pr_\rho g^J\big)}[]
satisfies
\eqn{\delta_\sigma Y_6^{\prime}\vphantom{Y}^\mu+6\hspace{1pt}\sigma\,
Y_6^{\prime}\vphantom{Y}^\mu=\E_6^\prime\vphantom{E}^{\mu\lambda}
\pr_\lambda\sigma}[]
for
\eqna{\E_6^\prime\vphantom{\E}^{\mu\lambda}= w_{KIJ}\big(&
2\,\pr^\rho g^K\nabla^\mu\pr^\lambda g^I\pr_\rho g^J
-\gamma^{\mu\lambda}\,\pr^\nu g^K\nabla^\rho\pr\nu g^I \pr_\rho g^J\\
{}& +4\,\pr^\rho g^K\nabla^{(\mu}\pr_\rho g^I\pr^{\lambda)}g^J
-2\, \pr^{(\mu}g^K\nabla^{\lambda)}\pr^\rho g^I\pr_\rho g^J\big)\, ,}[]
so long as $w_{KIJ}=-w_{KJI}, \, w_{IJK}+w_{JKI} + w_{KIJ}=0$.
This gives rise to an ambiguity in the last line of \Ysixsixd and
correspondingly the last line of $\E_6\vphantom{\E}^{\mu\lambda}$ in
\calYEsixd. In  \eqref{w24}  and \eqref{Ysixsixd} if $w_I = \pr_I u$ for any
scalar $u$ defined on the conformal manifold then the variation \eqref{Wvar}
can be removed by a local contribution to $W$. To obtain a monotonic RG flow
away from a critical point
it is necessary to add a term linear in $w_I\beta^I$ to $c, a$ when $d=2,4$.

Despite the  differences between six and two or four dimensions it is of
course possible that further assumptions may lead to relations between the
rank two tensors on the conformal manifold which could ensure that
$g_{1,IJ}$ is positive, at least in the neighbourhood of a fixed point, and
that there is then a potential perturbative $a$-theorem.  In particular
this might be the case in supersymmetric theories but also when a
nontrivial six dimensional CFT has a holographic dual. In such cases there
are arguments for an $a$-theorem which appear to be valid in any dimension
\cite{Freedman,Sinha}. Such arguments depend on positivity conditions for
the bulk energy momentum tensor which are doubtless vitiated in any
correspondence for $\phi^3$ theories. In any event, simple holographic
duals may not be sensitive to the additional two index tensors revealed by
our general discussion in six dimensions. Other arguments for a $c$, or
$a$, theorem in six dimensions are given in \cite{Yonekura}. This relates
the variation of the free energy on a sphere as the radius varies to the
metric defined by the two point function. A rather similar argument,
restricted to four dimensions, is given in \cite{Forte}.  The relation to
our analysis is not clear but the calculation is quite sensitive to the
details of regularisation.

\ack{We are both grateful for the warm hospitality of KITP Santa Barbara
where this work was initiated. We have benefitted from the
\emph{Mathematica} package \href{http://www.xact.es}{\texttt{xAct}}. HO
would like to thank Adam Schwimmer at KITP and Ian Jack for many helpful
discussions.  The research of AS is supported in part by the National
Science Foundation under Grant No.~1350180.}

\begin{appendix}

\section{Conformal Tensors, Invariants and Operators}
\label{conformal}

The anomalous terms in Weyl scaling identities are, for type B \cite{Deser},
expressed in terms of conformal scalars. These are in turn
formed from conformal tensors which transform homogeneously, without
any derivatives of $\sigma$. Concise expressions for these may be obtained
by first defining a modified scalar curvature ${\hat R}$ (in the mathematical
literature this is commonly denoted by $J$) and the Schouten tensor $P_{\mu\nu}$
given by
\be
{\hat R} = \frac{1}{2(d-1)} \, R \, , \quad P_{\mu\nu} =
\frac{1}{d-2} ( R_{\mu\nu} - \gamma_{\mu\nu} {\hat R} ) \, , \qquad
\gamma^{\mu\nu} P_{\mu\nu} = {\hat R}\, ,  \quad \nabla^\nu P_{\mu\nu}
= \pr_\mu {\hat R}\, .
\label{defPR}
\ee
These have the crucial properties under Weyl rescalings of the metric
\be
\delta_\sigma {\hat R}  = - 2\hspace{1pt} \sigma \, {\hat R}  - \nabla^2 \sigma \, ,
\qquad
\delta_\sigma  P_{\mu\nu} = - \nabla_\mu \pr_\nu \sigma \, .
\label{PRsig}
\ee
The  Weyl tensor is then given in terms of the Riemann tensor by
\be
W_{\lambda\rho\mu\nu} =  R_{\lambda\rho\mu\nu} - \gamma_{\lambda\mu} \, P_{\rho\nu}
+ \gamma_{\rho\mu} \, P_{\lambda\nu} + \gamma_{\lambda\nu} \, P_{\rho\mu} -
\gamma_{\rho\nu} \, P_{\lambda\mu} \, .
\ee
To discuss tensors which transform homogeneously under Weyl  rescaling
it is necessary to consider the Cotton tensor defined by
\be
C_{\mu\nu\lambda} = \nabla_\lambda P_{\mu\nu} - \nabla_\nu P_{\mu\lambda}\,
,
\ee
and also the Bach tensor given by
\eqna{B_{\mu\nu} ={}&  \nabla^\lambda C_{\mu\nu\lambda}
- P^{\lambda\omega} W_{\lambda\mu\nu\omega} \\
={}& - 2 \, P^{\lambda\omega} W_{\lambda\mu\nu\omega} - d \,
P_{\mu\lambda}P^\lambda{}_\nu + \gamma_{\mu\nu} \, P_{\rho\lambda}P^{\rho\lambda}
+ \nabla^2 P_{\mu\nu} - \nabla_\mu \nabla_\nu {\hat R} \, .}[]
These have the properties
\eqna{& C_{\mu\nu\lambda}=-C_{\mu\lambda\nu}\, , \quad C_{\mu\nu\lambda} +
C_{\lambda\mu\nu}+ C_{\nu\lambda\mu}= 0 \, ,\quad  \gamma^{\mu\nu} C_{\mu\nu\lambda}=0 \, , \quad
\nabla^\mu C_{\mu\nu\lambda}=0 \, , \\
& B_{\mu\nu} = B_{\nu\mu} \, , \qquad  \gamma^{\mu\nu}B_{\mu\nu}=0 \, , \qquad
\nabla^\nu B_{\mu\nu} = (d-4) \, P^{\lambda\rho}C_{\lambda\rho\mu} \, .}[]
The Bianchi identity for the Weyl tensor
becomes
\eqna{& \nabla_\omega W_{\lambda\rho\mu\nu} + \nabla_\mu W_{\lambda\rho\nu\omega} +
\nabla_\nu W_{\lambda\rho\omega\mu} \\
&{} = \gamma_{\lambda\mu}\, C_{\rho \omega\nu} +  \gamma_{\rho\mu}\, C_{\lambda\nu \omega}
+  \gamma_{\lambda\nu}\, C_{\rho \mu\omega} +  \gamma_{\rho\nu}\, C_{\lambda \omega\mu}
+ \gamma_{\lambda\omega}\, C_{\rho \nu\mu} +  \gamma_{\rho\omega}\, C_{\lambda\mu\nu} \, ,}[bianchi]
from which  $\nabla^\rho W_{\rho\mu\nu\lambda} = - (d-3) \, C_{\mu\nu\lambda}$.
Under Weyl scalings $\delta_\sigma W_{\lambda\rho\mu\nu} = 2\hspace{1pt} \sigma \,
W_{\lambda\rho\mu\nu}$ and
\be
\delta_\sigma C_{\mu\nu\lambda} = - \pr^\rho \sigma \,
W_{\rho\mu\nu\lambda} \, , \qquad
\delta_\sigma B_{\mu\nu} = - 2\hspace{1pt}\sigma B_{\mu\nu}  + (d-4) \,
\pr^\lambda \sigma ( C_{\mu\nu\lambda} + C_{\nu\mu\lambda}) \, .
\ee
Since the Weyl  tensor vanishes when $d=3$ the Cotton tensor is then a conformal
tensor, as is the Bach tensor when $d=4$.

In terms of these expressions
\be
E_4 = 6 \, R_{\lambda\rho}{}^{[\lambda\rho}R_{\mu\nu}{\, }^{\mu\nu]}
= W_{\lambda\rho}{}^{\mu\nu} W_{\mu\nu}{}^{\lambda\rho} - 4(d-2)(d-3)
(P^{\mu\nu} P_{\mu\nu} - {\hat R}^2 ) \, ,
\label{Efour}
\ee
which is the Euler density in four dimensions,
and also
\eqna{E_6 = {}& 90 \, R_{\lambda\rho}{}^{[\lambda\rho}R_{\mu\nu}{\, }^{\mu\nu}
R_{\omega\tau}{}^{\omega\tau]} \\
={}&8 \,I_1 +4 \, I_2 + 6(d-5)\big ( {\hat R}\,
W_{\lambda\rho}{}^{\mu\nu} W_{\mu\nu}{}^{\lambda\rho}
- 4 \, P_{\mu\nu} \, W^{\mu\lambda\rho\omega} W^\nu{}_{\lambda\rho\omega} \big ) \\
&{} - 24(d-4)(d-5) \, P_{\mu\nu}P_{\lambda\rho}\,  W^{\mu\lambda\rho\nu} \\
&{}+  8(d-3)(d-4)(d-5) \big ( {\hat R}^3 -3\, {\hat R}\,  P^{\mu\nu} P_{\mu\nu}
+ 2\, P^{\mu\nu} P_{\nu\lambda}P^{\lambda}{}_\mu \big )  \, ,}[Esix]
for $I_1,I_2$ conformal scalars
\be
I_1 =  W_{\rho\mu\nu\lambda}\,W^{\mu\omega\tau\nu}\, W_\omega{}^{\rho\lambda}{}_\tau \, ,
\qquad
I_2 = W_{\mu\nu}{}^{\lambda\rho}\, W_{\lambda\rho}{}^{\omega\tau} \,
W_{\omega\tau}{\,}^{\mu\nu} \, .
\label{scalar6}
\ee
These  satisfy
\eqna{\delta_\sigma E_4 + 4\hspace{1pt} \sigma E_4 =  {}& 8(d-3)\,
\nabla_\mu ( G_4\vphantom{G}^{\mu\nu} \pr_\nu \sigma) \, , \\
\delta_\sigma E_6 + 6\hspace{1pt} \sigma E_6 =  {}& 24(d-5)\,
\nabla_\mu ( G_6\vphantom{G}^{\mu\nu} \pr_\nu \sigma) \, ,}[]
for
\be
G_4\vphantom{G}^{\mu\nu} = ( d-2) ( P^{\mu\nu} - \gamma^{\mu\nu} \, {\hat R} )
= R^{\mu\nu} - \half \gamma^{\mu \nu} R  \, ,
\label{Gfour}
\ee
the Einstein tensor, and
\eqna{G_6\vphantom{G}^{\mu\nu} = {}& W^{\mu\lambda\rho\omega} W^\nu{}_{\lambda\rho\omega}
+ 2(d-4) \, W^{\mu\lambda\rho\nu} P_{\lambda\rho} - 2(d-3)(d-4)
( P^{\mu\lambda}P^\nu{}_\lambda - P^{\mu\nu} \, {\hat R} ) \\
&{} - \tfrac{1}{4} \gamma^{\mu\nu} \big ( W^{\tau\lambda\rho\omega}
W_{\tau\lambda\rho\omega} - 4 (d-3)(d-4) ( P^{\lambda\rho}P_{\lambda\rho}
- {\hat R}^2 ) \big )\, ,}[]
where $\nabla_\mu G_4\vphantom{G}^{\mu\nu} = \nabla_\mu G_6\vphantom{G}^{\mu\nu} = 0$
and $G_6\vphantom{G}^{\mu\nu} = 0$ for $d=3,4$. For completeness we note that
\be
E_2 = R \, ,  \qquad \delta_\sigma E_2 + 2\hspace{1pt} \sigma E_2 =  2(d-1)\,
\nabla_\mu ( G_2\vphantom{G}^{\mu\nu} \pr_\nu \sigma) \, ,  \quad \mbox{for}
\quad G_2\vphantom{G}^{\mu\nu} = - \gamma^{\mu\nu} \, .
\ee

It is useful to note that
\be
 \delta_{\sigma} G_4\vphantom{G}^{\mu\nu} + 4\hspace{1pt}\sigma\, G_4\vphantom{G}^{\mu\nu}
 = - (d-2)\,  H_4\vphantom{H}^{\mu\lambda\rho\nu} \, \nabla_\lambda \pr_\rho \sigma \, , \quad
 H_4\vphantom{H}^{\mu\lambda\rho\nu} =
  \gamma^{\mu\rho} \gamma^{\lambda\nu}  - \gamma^{\lambda\rho} \gamma^{\mu\nu} \, ,
\label{G4var}
\ee
and
\be
 \delta_{\sigma} G_6\vphantom{G}^{\mu\nu} + 6\hspace{1pt} \sigma  \, G_6\vphantom{G}^{\mu\nu} = -2 (d-4)\,
 H_6\vphantom{H}^{\mu\lambda\rho\nu}
\, \nabla_\lambda \pr_\rho \sigma \, ,
\ee
for
 \begin{align}
   H_6\vphantom{H}^{\mu\lambda\rho\nu} = {}& W^{\mu\lambda\rho\nu}\nn \\
\noalign{\vskip - 5pt}
& {} - (d-3) \big ( \gamma^{\mu\rho} P^{\lambda\nu}
- \gamma^{\lambda\rho} P^{\mu\nu} -  \gamma^{\mu\nu} P^{\lambda\rho}
+  \gamma^{\lambda\nu} P^{\mu\rho} - {\hat R}
( \gamma^{\mu\rho} \gamma^{\lambda\nu}  - \gamma^{\lambda\rho} \gamma^{\mu\nu} ) \big ) \, ,
\label{Hsix}
\end{align}
where $\nabla_\rho H_6{\vphantom{H}}^{\mu\lambda\rho\nu}= 0, \,
\gamma_{\lambda\rho} H_6{\vphantom{H}}^{\mu\lambda\rho\nu} = (d-3) \,
G_4\vphantom{G}^{\mu\nu}$.

Besides $I_1,I_2$ in \eqref{scalar6} there is an additional conformal scalar of
dimension six. For general $d$ it may be succinctly expressed as
\eqna{\Omega = {}&
\tfrac{1}{4}(10-d) \big (
W^{\rho\mu\nu\lambda} \nabla^2  W_{\rho\mu\nu\lambda}
+ 4(d-2)\, C^{\mu\nu\lambda}C_{\mu\nu\lambda}\big ) \\
&{}  + \big (\tfrac{1}{8}(d-2) \,  \nabla^2
- 4 \, {\hat R} \big )W^{\rho\mu\nu\lambda}W_{\rho\mu\nu\lambda} \, .}[om6]
Alternative forms \cite{Feff}, \cite{Parker}, \cite{Erdmenger}, equivalent to
\eqref{Ithree} up to contributions linear in $I_1,I_2$, can be obtained with
the aid of the relations from \eqref{bianchi}
\eqna{4\,  I_1 - I_2 = {}&  W^{\rho\mu\nu\lambda} \nabla^2 W_{\rho\mu\nu\lambda}
- 2(d-2)\,  P_{\rho\omega} \, W^{\rho\mu\nu\lambda}\, W^\omega{}_{\mu\nu\lambda}
- 2 \, {\hat R} \,  W^{\rho\mu\nu\lambda}W_{\rho\mu\nu\lambda} \\
&{} + 2(d-2)(d-3) \, C^{\mu\nu\lambda} C_{\mu\nu\lambda}
+ 2(d-2)\,  \nabla_\omega  (W^{\omega \mu\nu\lambda} C_{\mu\nu\lambda}  )\, , \\
 (d-4) \nabla_\omega  (& W^{\omega \mu\nu\lambda}  C_{\mu\nu\lambda}  ) =
- \nabla_\mu \nabla_\nu ( W^{\mu \lambda \rho \omega} W^\nu{}_{\lambda\rho\omega})
+ \tfrac{1}{4} \,\nabla^2 ( W^{\mu \lambda \rho \omega} W_{\mu\lambda\rho\omega})\, .}[deriv1]
The form used in \cite{Bastianelli} is given by
\be
I_3 =  (d-3) \Omega -  \tfrac{1}{2}(10-d) (4I_1 - I_2) \, ,
\ee
so that, for $d=6$,
\eqna{I_3={}& W^{\rho\mu\nu\lambda} \nabla^2  W_{\rho\mu\nu\lambda} + 16 \,
P_{\mu\nu} \, W^{\mu\rho\lambda\omega}\,  W^\nu{}_{\rho\lambda\omega}
- 8 \, {\hat R} \,  W^{\rho\mu\nu\lambda} \, W_{\rho \mu\nu \lambda} \\
 &{}+ 8 \, \nabla_\mu \nabla_\nu ( W^{\mu \lambda \rho \omega}
W^\nu{}_{\lambda\rho\omega})
 - \half \, \nabla^2 (W^{\rho\mu\nu\lambda} \, W_{\rho \mu\nu \lambda} )\,
 .}[Ithree]
The $I_r$ all satisfy
\be
\delta_\sigma I_r + 6\hspace{1pt} \sigma\, I_r =0 \, .
\ee
Besides \eqref{deriv1} we may also note the derivative relation
\eqna{\nabla_\mu \nabla_\nu  \big ( P^{\mu\lambda}P^\nu{}_\lambda {}
- 2\, P^{\mu\nu} {\hat R} \big ) + \nabla^2 {\hat R}^2
={}&  P_{\mu\nu}P_{\lambda\rho}\,  W^{\mu\lambda\rho\nu} + d\,
P^{\mu\nu} P_{\nu\lambda}P^{\lambda}{}_\mu - {\hat R} \, P^{\mu\nu}P_{\mu\nu}  \\
&{} - \half \, C^{\lambda\mu\nu}C_{\lambda\mu\nu} + \nabla^\lambda P^{\mu\nu}
\nabla_\lambda P_{\mu\nu} - \pr^\lambda {\hat R} \, \pr_\lambda {\hat R} \, .
}[deriv2]

If a connection $A_\mu$, with corresponding field strength $F_{\mu\nu}$, is
present, then there are further conformal scalars. Analogous to \eqref{om6}
there is a similar dimension six conformal scalar formed from $F_{\mu\nu}$
which as given in \cite{Parker} has the form
\eqna{{\hat \Omega} = {}& \tfrac14 (10-d) \big (\quar (d-4) \,(
F^{\mu\nu} \D^2 F_{\mu\nu} + \D^2 F^{\mu\nu} \, F_{\mu\nu} )
+ \D_\mu F^{\mu \lambda} \, \D^\nu F_{\nu \lambda} \big ) \\
&{}+ \tfrac{1}{16}(d-4) \big ( (d-4) \D^2 - 24 \, {\hat R} \big ) F^{\mu\nu}
F_{\mu\nu} \, ,}[]
for $\D_\mu$ the appropriate covariant derivative, $\D_\lambda F_{\mu\nu}
= \pr_\lambda F_{\mu\nu} +  [A_\lambda ,F_{\mu\nu}]$. Corresponding to
\eqref{deriv1}, using the Bianchi identity for $F_{\mu\nu}$,
\eqna{\D_\mu \D_\nu & ( F^{\mu\lambda} F^\nu{}_\lambda ) -
\half \, \D^2(F^{\mu\nu} F_{\mu\nu}) \\
= {}& \D_\mu F^{\mu \lambda} \, \D^\nu F_{\nu \lambda} - \half \,
\D^\lambda F^{\mu\nu} \, \D_\lambda F_{\mu\nu}
- (d-4) \, P_{\mu\nu} F^{\mu\lambda} F^\nu{}_\lambda - {\hat R} \,
F^{\mu\nu} F_{\mu\nu} \\
&{}+ \half \, W_{\mu\nu\lambda\rho}\, F^{\mu\nu}F^{\lambda\rho} - 2 \,
F^{\mu\nu}F_{\nu\lambda}F^\lambda{}_\mu \, .}[deriv3]
The terms in the last line are conformal scalars. Using \eqref{deriv3}
an expression similar to \eqref{Ithree} can be obtained which is more
convenient for our purposes. For $d=6$ this becomes
\begin{align}
{\hat I} = {}& \tfrac12 (
F^{\mu\nu} \D^2 F_{\mu\nu} + \D^2 F^{\mu\nu} \, F_{\mu\nu} )
 - 4\, {\hat R}  \, F^{\mu\nu} F_{\mu\nu}
+ ( 2 \, \D_\mu \D_\nu   + 4 \, P_{\mu\nu} ) (F^{\mu\lambda}F^\nu{\!}_\lambda )
 \, ,
\label{Finv}
\end{align}
which corresponds to the form given in \eqref{LsixF}.

In addition to conformal tensors there are also conformally covariant
differential operators\foot{An overview and some useful expressions can
be found in \cite{Gover}.} which play a crucial role. The conformal Laplacian,
or Yamabe operator,
\be \Delta_2 = - \nabla^2 + \half (d-2) \,{\hat R} \, ,
\label{del2}
\ee
acts on scalars of dimension $\half (d-2)$, $\delta_\sigma
\Delta_2 = - \half (d+2) \, \sigma \,
\Delta_2 + \half (d-2) \, \Delta_2 \, \sigma$. For $d=10$ the conformal scalar
$\Omega$ in \eqref{om6} is just
$-\Delta_2 \, W^{\rho\mu\nu\lambda} \, W_{\rho \mu\nu \lambda}$. The
corresponding fourth order Paneitz operator \cite{Paneitz} was for $d=4$
found  first by Fradkin and Tseytlin \cite{Fradkin} and also rederived by
Riegert \cite{Riegert},
\eqna{
\Delta_4 = {}& \Delta_{4,1} + \half (d-4) \, Q_4 \, , \qquad
 \Delta_{4,1} = \nabla^2 \nabla^2 + \nabla^\mu \big ( 4  \, P_{\mu \nu}
- (d - 2)\, \gamma_{\mu\nu} {\hat R} \, \big ) \pr^\nu \, ,  \\
Q_4 = {}& - \nabla^2 {\hat R} - 2 \, P^{\mu\nu}P_{\mu\nu}
+ \half d \, {\hat R}^2 \, , \qquad \delta_\sigma Q_4 + 4 \, \sigma\, Q_4
=  \Delta_{4,1} \sigma \, .
}[]
$\Delta_4$ acts
on scalars such that $\delta_\sigma \Delta_4 = - \half(d+4)
\hspace{1pt} \sigma \,\Delta_4 + \half(d-4) \, \Delta_4 \, \sigma$.
For $d=4$ this expression for $ \Delta_{4,1}$ is equivalent to ithe second
line in the result for $L_4$ in
\eqref{Lfour}. There is a corresponding extension to $(\nabla^2)^3$, first constructed
by Branson \cite{Branson}, which can be written as
\be
\Delta_6 =  \Delta_{6,1} + \half (d-6) \, Q_6 \, ,
\ee
where
\eqna{
\Delta_{6,1} ={}&  -\nabla^2 \nabla^2 \nabla^2
- 8 \big (  \nabla^2 P_{\mu\nu}\nabla^\mu\pr^\nu
+ \nabla^\mu \nabla^\nu  P_{\mu \nu} \nabla^2 \big ) + \tfrac32(d-2)  \, \nabla^2
{\hat R}\,  \nabla^2 \\
& - 2 \, \nabla^\mu \bigg ( (10-d) \, \nabla_\mu \nabla_\nu {\hat R}
+ \frac{8}{d-4} \, B_{\mu\nu}
  + 24 \, P_{\mu\lambda}P_\nu{}^\lambda
- 4(d-2) \, P_{\mu\nu}{\hat R} \bigg ) \pr^\nu \\
&{}+ 4\, \nabla^\mu  \big (  \nabla^2 {\hat R}  + (d-4) \, P_{\rho\lambda}
P^{\rho\lambda}  - (\tfrac{3}{16}(d-2)^2  -1) \, {\hat R}^2 \big )\pr_\mu \, ,
}[Del6]
so that $\delta_\sigma \Delta_6 = - \half(d+6) \hspace{1pt} \sigma \, \Delta_6
+  \half(d-6) \hspace{1pt} \Delta_6 \, \sigma $. For $d=6$, $\Delta_{6,1}$ is
equivalent to the contributions $S_1+S_2+S_3$ as given by \eqref{S1}, \eqref{S2},
\eqref{S3}. The expression for $\Delta_{6,1}$ in \eqref{Del6}
was obtained by seeking a Weyl invariant
$S=  \int \rmd^d x \sqrt{-\gamma} \; \big ( \half \, \pr^\mu \nabla^2 \vphi \, \pr_\mu \nabla^2\vphi
+ \cdots \big )$, assuming $\delta_\sigma \vphi = -\half (d-6)\sigma \,  \vphi$, in a similar fashion
 to the discussion in section 2.
$Q_6$ is determined by
\be
\delta_\sigma Q_6 + 6 \, \sigma \, Q_6 = \Delta_{6,1}\sigma \quad \mbox{or} \quad
 \delta_\sigma Q_6 + \half(d+6) \, \sigma \, Q_6 = \Delta_{6} \sigma\, .
\ee
It is easy to verify integrability $[\delta_\sigma, \delta_{\sigma'} ] Q_6 = 0$.
A minimal solution is given by
\eqna{
Q_6 = {}& \nabla^2 \nabla^2 {\hat R} + \nabla^2 \big ( 4 \, P^{\mu\nu}P_{\mu\nu}
- \half(d-6) \, {\hat R}^2 \big ) + 8 \, P^{\mu\nu}\, \nabla_\mu \pr_\nu {\hat R}
- \half(d+10) \, {\hat R} \, \nabla^2 {\hat R} \\
&{}+ 16\, P^{\mu\nu} P_{\nu\lambda}P^{\lambda}{}_\mu + \frac{16}{d-4} \,
B^{\mu\nu} P_{\mu\nu} - 4d \, P^{\mu\nu}P_{\mu\nu} \, {\hat R}
+ \tfrac14 (d-2)(d+2) \, {\hat R}^3 \, .
}[Q6]
A non zero  Bach tensor is clearly an obstruction to the operator existing for $d=4$.
The expression for the Branson operator can be written in various different
forms; that given by \eqref{Del6} and \eqref{Q6} appears simpler than most.

Besides acting on scalars there are also conformal differential operators
for tensors with various symmetries. For our purposes we need only consider
operators acting on symmetric traceless tensors of rank two. Adapting
results from \cite{wunsch, ErdmengerO} to this special case
\eqna{\Delta_{2,T} \, h_{\mu\nu} = {}& \Delta_2 \,h_{\mu \nu} + \frac{8}{d+2} \, \nabla_{(\mu}
\nabla^\lambda h_{\nu)\lambda} + 4 \, P_{(\mu}{}^\lambda h_{\nu)\lambda} \\
&{} - \frac{1}{d} \, \gamma_{\mu\nu} \Big (  \frac{8}{d+2} \, \nabla^\rho
\nabla^\lambda h_{\rho\lambda} + P^{\rho\lambda} h_{\lambda\rho} \Big)}[]
so that
$\delta_\sigma \Delta_{2,T} = \half (d-6) \, \Delta_{2,T} \, \sigma - \half (d-2) \, \sigma \,
\Delta_{2,T}$. The operators $\Delta_2$ and $\Delta_{2,T}$ are implicitly determined
by the $j_2,j_1$ contributions in $T_1+T_2$ given by \eqref{T1}, \eqref{T2}.

The calculations for $\phi^3$ theory are based on using the heat kernel expansion
for $e^{- t \,  \Delta} $, with $\Delta = - \D^2 +\half (d-2) {\hat R} + Y
$  in terms of the Seeley--DeWitt
coefficients $a_n(x,y)$.  $\Delta$ is a conformal differential operator if we assume
$\delta_\sigma Y = - 2\hspace{1pt}\sigma \, Y$. If $A_\mu,Y=0$ then   $ \Delta = \Delta_2$ as in \eqref{del2}.
For the diagonal coefficients $a_n |$, when $y=x$, we have
\eqna{180\, a_2| ={}& W^{\lambda\rho\mu\nu} W_{\lambda\rho\mu\nu} +
15 \, F^{\mu\nu}F_{\mu\nu} + 60 \, Y^2 + 30\,
\Delta Y \\
&{} -(d-6) \big ( (d-2) P^{\mu\nu} P_{\mu\nu}
- \half (5d-16) {\hat R}^2 + 3 \nabla^2 {\hat R}  - 15 \, {\hat R}\, Y \big )  \\
\noalign{\vskip 2pt}
={}& \tfrac{3}{2} \,  W^{\lambda\rho\mu\nu} W_{\lambda\rho\mu\nu} - \half \, E_4
+ 15 \, F^{\mu\nu}F_{\mu\nu} + 90\, Y^2 +  6 \, \nabla^2 {\hat R}
- 30 \, \D^2 Y \quad \mbox{if} \quad d=4 \, ,}[afour]
and from \cite{Gilkey} for $A_\mu,Y=0$,
\eqna{
7!\, a_3| ={}& - \tfrac{80}{9} \, I_1 + \tfrac{44}{9}\,  I_2 + 6\, \Omega \\
&{} + (d-8) \big ( -
\tfrac{3}{2} \,  \nabla^\omega W^{\rho\mu\nu\lambda} \nabla_\omega  W_{\rho\mu\nu\lambda}
- \tfrac{16}{3} \, P_{\rho\omega} \, W^{\rho\mu\nu\lambda}\,  W^\omega{}_{\mu\nu\lambda}
- \tfrac{14}{3} \, {\hat R}  \, W^{\rho\mu\nu\lambda}\,  W_{\rho\mu\nu\lambda}  \\
& \hskip 1.8cm {}  +  \tfrac{8}{3} (d+2) \, P_{\mu\nu}P_{\lambda\rho}\,  W^{\mu\lambda\rho\nu}
+ 8(d-2)\,  C^{\mu\nu\lambda}C_{\mu\nu\lambda} \\
& \hskip 1.8cm {}  + (d-2)\big (2\, \nabla^\lambda P^{\mu\nu} \nabla_\lambda P_{\mu\nu} - 4 \,
\nabla^2 (P^{\mu\nu} P_{\mu\nu} ) - 4 \, \nabla_\mu\nabla_\nu ( P^{\mu\nu} {\hat R} )\big  ) \\
& \hskip 1.8cm {} - (5 d - 22) \, \pr^\lambda {\hat R} \, \pr_\lambda {\hat R} + (9d-  32 ) \,
\nabla^2 ({\hat R}^2 ) - 6 \, \nabla^2 \nabla^2 {\hat R} \\
& \hskip 1.8cm {} +\tfrac{8}{9} ( d^2 - 4d +12) \, P^{\mu\nu} P_{\nu\lambda}P^{\lambda}{}_\mu
+\tfrac{2}{3}(7d^2 - 40  d +36)  \, {\hat R}\, P^{\mu\nu}P_{\mu\nu} \\
& \hskip 1.8cm {} - \tfrac{1}{9} ( 35d^2 - 266 d + 456 ) \, {\hat R}^3 \big
) \, .}[asix]
For $d=6$
\eqna{
7!\, a_3| ={}& \tfrac{5}{9} \, E_6 - \tfrac{28}{3} \, I_1
+ \tfrac{5}{3} \, I_2 + 2\, I_3 + 14\big (3\, {\hat I}
+ 5\, W_{\mu\nu\lambda\rho}\, F^{\mu\nu}F^{\lambda\rho} -8 \,
F^{\mu\nu}F_{\nu\lambda}F^\lambda{}_\mu \big )  \\
&{} - 7! \big ( \tfrac{1}{12} \, Y^3 +
\tfrac{1}{12} \, Y \Delta Y + \tfrac{1}{180} \,
W^{\rho\mu\nu\lambda}  W_{\rho\mu\nu\lambda}\, Y \big ) \\
&{}  - 7! \big(
\tfrac{1}{30 }\,(  F^{\mu\nu}F_{\mu\nu} \,Y + Y\, F^{\mu\nu}F_{\mu\nu} )
+ \tfrac{1}{60} \, F^{\mu\nu} \, Y F_{\mu\nu} \big ) \\
&{} - \nabla_\mu \nabla_\nu
\big ( 12 \, W^{\mu\lambda\rho\omega} W^\nu{\!}_{\lambda\rho\omega}
+ 16 \, P^{\mu\lambda}P^\nu{\!}_\lambda - 64 \, P^{\mu\nu} {\hat R}\big ) \\
&{}+ \nabla^2 \big (
\tfrac{9}{2} \, W^{\lambda\rho\mu\nu} W_{\lambda\rho\mu\nu} + 32 \,
P^{\mu\nu}P_{\mu\nu} - 60\, {\hat R}^2 \big ) +12 \, \nabla^2 \nabla^2 {\hat R} \\
&{}- 56 \, \D_\mu \D_\nu (F^{\mu\lambda}F^\nu{\!}_{\lambda}) + 49 \,
\D^2 (F^{\mu\nu}F_{\mu\nu} ) \\
&{} - 7! \big (  \tfrac{1}{90} \, \D_\mu \D_\nu ( G_4\vphantom{G}^{\mu\nu} Y) -\tfrac{1}{24} \, \D^2 Y^2
+ \tfrac{1}{60} \, \D^2 \D^2 Y \big )  \, .}[asix6d]
This gives the results in \eqref{athree} and \eqref{oneloop}.
The results in \eqref{afour}, \eqref{asix} and \eqref{asix6d} reflect the
theorems of Parker and Rosenberg \cite{Parker}\footnote{As noted in
\cite{Bastianelli} their results contain some  errors which are hopefully
corrected in \eqref{asix}, \eqref{asix6d}.}
that $a_{n}|$ for $d=2n+2$ is a conformal scalar and for $d=2n$,
$\int \rmd^{2n} x \sqrt{-\gamma} \; a_n| $ is a conformal invariant, and
suggest the slight extension, that for $d=2n$,  $a_n|$ is a linear
combination of conformal scalars and the Euler density $E_{2n}$ up to
terms with two derivatives.

\newsec{Expansion of Six Dimensional Dilaton Action}[sixddilaton]

In six dimensions $\cL_6(\sigma)$ in \eqref{wsig}
may be obtained by using \eqref{Xrecur}. Starting from \eqref{Xsix} we  may
straightforwardly use \eqref{PRsig} successively in \eqref{Xrecur}
to determine $X_{6,r}^{R\, \mu\nu}$ for $r=1,2,3,4$ and hence obtain
\eqna{\cL_6^R(\sigma) = {}& \sigma \, L_6^R  - 12 \,
a \, G_6\vphantom{G}^{\mu\nu} \pr_\mu \sigma \pr_\nu \sigma \\
&{}+ 16\, a \big ( W^{\mu\lambda\rho\nu}\,  \nabla_\lambda
\pr_\rho \sigma\, \pr_\mu \sigma \pr_\nu\sigma  \,
-  6\, P^{\mu\nu} \, \nabla^\lambda \pr_\mu \sigma\,
\pr_\nu \sigma \pr_\lambda \sigma \\
&\hskip 1.3cm {} + 3 \, P^{\mu\nu}\, \nabla^2 \sigma \,  \nabla_\mu \pr_\nu \sigma
+ 3 \, P^{\mu\nu} \, \nabla_\mu \pr_\nu \sigma \,
\pr^\lambda \sigma \pr_\lambda \sigma
\pr_\lambda \sigma \\
&\hskip 1.3cm {} + 3 \,  {\hat R}\,
\nabla^\mu \pr^\nu\sigma \, \pr_\mu \sigma \pr_\nu \sigma
-  3 \,  {\hat R}\,  \nabla^2 \sigma \,
\pr^\lambda \sigma \pr_\lambda  \sigma \big ) \\
&{} - 24\, a \big (  \tfrac{5}{2} \, {\hat R}\,
(\pr^\lambda \sigma \pr_\lambda \sigma )^2
+ \nabla^\mu\pr^\nu \sigma \nabla_\mu \pr_\nu \sigma \,
\pr^\lambda \sigma \pr_\lambda \sigma
- (\nabla^2 \sigma)^2  \, \pr^\lambda \sigma \pr_\lambda \sigma \big )  \\
&{} + 36 \, a \, \nabla^2 \sigma \, (\pr^\lambda \sigma \pr_\lambda \sigma)^2
+ 24\, a \,  ( \pr^\lambda \sigma \pr_\lambda \sigma)^3 \, ,}[]
which matches \cite{Elvang}. For the contributions arising from $L_6^g$
given by \eqref{L6g} and $X_6^{g\, \mu\nu}$ given by \eqref{X6g}
\eqna{\cL_6^g(\sigma) = {}& \sigma \, L_6^g  + \half \,
X_6^{g\, \mu\nu} \pr_\mu \sigma \pr_\nu \sigma \\
&{} + 2\, g_{IJ} \pr^\mu g^I \pr^\nu g^J \big ( 6\, \nabla^\lambda \pr_\mu \sigma \,
\pr_\nu \sigma \pr_\lambda \sigma +\nabla_\mu \pr_\nu \sigma \, \pr^\lambda \sigma
\pr_\lambda \sigma - 2\, \nabla^2 \sigma \, \pr_\mu \sigma \pr_\nu \sigma \\
\noalign{\vskip -1pt}
&\hskip 3cm {}- 4 \, \pr_\mu \sigma \pr_\nu \sigma\,
\pr^\lambda \sigma \pr_\lambda \sigma \big ) \\
&{}-  g_{IJ} \pr^\lambda g^I \pr_\lambda g^J \big ( 6 \, \nabla^\mu \pr^\nu \sigma \,
\pr_\mu \sigma \pr_\nu \sigma + \nabla^2 \sigma \, \pr^\mu \sigma \pr_\mu \sigma
- (\pr^\mu \sigma \pr_\mu \sigma)^2 \big ) \, .}[]
The remaining contributions from \eqref{L6j} with \eqref{X6j} and \eqref{X6k}
are then
\begin{align}
\cL_6^j(\sigma) =  \sigma \, L_6^j  - \half \,
X_6^{j\, \mu\nu} \pr_\mu \sigma \pr_\nu \sigma \, ,  \qquad
\cL_6^k(\sigma) =  \sigma \, L_6^k \, .
\end{align}

\newsec{Fermions}[resFermions]

For completeness we extend the results in \cite{Bastianelli} to include
background gauge fields coupled to fermion conserved currents
${\bar \psi} \gamma^\mu t_a \psi$. In this case the
one loop action is determined by an operator $\Delta = - {\slashed \D}^2$,
with $\D_\mu$ including the spinor and gauge connections.
This can be reduced to the form \eqref{deltan} where
\be
F_{\mu\nu} \to \quar \, R_{\mu\nu\lambda\rho} \, \gamma^\lambda \gamma^\rho
+ F_{\mu\nu}  \, 1_S\, ,
\qquad Y \to \half \, {\hat R} \, 1_S - \half \, F_{\mu\nu} \gamma^\mu \gamma^\nu \, ,
\label{spin}
\ee
with $1_S$ the spinor identity. For fermions then
\be
L_6^R + L_6^F = - \tr \big (a_{\Delta,3}| \big )
+ \nabla_\mu \nabla_\nu Z^{\mu\nu} \, ,
\label{LRF}
\ee
where the trace is over both spinorial and gauge indices.
In the formula \eqref{athree} for $\tr(a_{\Delta,3}| )$ we may use
\eqref{spin} to obtain in six dimensions, using $\tr(1_S) =8$,
\eqna{\tr({\hat I}) \to {}& \tfrac{1}{3} ( 4 I_1 - I_2 - 4 I_3 ) + 20\,
P_{\mu\nu} W^{\mu\lambda \rho \omega} W^\nu{}_{\lambda\rho\omega}
- 6 \, {\hat R} \, W^{\mu\nu\lambda\rho}W_{\mu\nu\lambda\rho}  \\
&{} - 20 \, {\hat R} \, \nabla^2 {\hat R} - 112 \,
 P^{\mu\nu}P_{\nu\lambda}P^{\lambda}{}_\mu + 56\,  {\hat R} \,
P^{\mu\nu} P_{\mu\nu} + 16 \,  {\hat R}^3   \\
&{}+8\,  \tr({\hat I}) \, ,  \\
\noalign{\vskip 2pt}
\tr ( F^{\mu\nu} F_{\nu \lambda} F^\lambda{}_\mu) \to {}&
- I_1 - 3 \, P_{\mu\nu} W^{\mu\lambda \rho \omega} W^\nu{}_{\lambda\rho\omega}
- 6 \, P_{\mu\nu} P_{\lambda\rho} W^{\mu\lambda\rho\nu}   \\
&{}- 20 \, P^{\mu\nu}P_{\nu\lambda}P^{\lambda}{}_\mu + 18 \, {\hat R} \,
P^{\mu\nu} P_{\mu\nu} + 2 \, {\hat R}^3  \\
&{} + 8 \, \tr ( F^{\mu\nu} F_{\nu \lambda} F^\lambda{}_\mu) \, ,  \\
\noalign{\vskip 2pt}
W_{\mu\nu\lambda\rho} \, \tr(F^{\mu\nu} F^{\lambda\rho}) \to {}&  - I_2
-  8 \, P_{\mu\nu} W^{\mu\lambda \rho \omega} W^\nu{}_{\lambda\rho\omega}
+ 8\, P_{\mu\nu} P_{\lambda\rho} W^{\mu\lambda\rho\nu}  \\
&{}+ 8\, W_{\mu\nu\lambda\rho} \, \tr(F^{\mu\nu} F^{\lambda\rho})  \\
\noalign{\vskip 2pt}
\tr ( F^{\mu\nu}F_{\mu\nu} \, Y ) \to &{} -\half  \big ( {\hat R}\,
 W^{\mu\nu\lambda\rho}W_{\mu\nu\lambda\rho}
+  16 \, {\hat R}\,  P^{\mu\nu}P_{\mu\nu}  + 4 \, {\hat R}^3 \big  )  \\
&{}+ 4\, W_{\mu\nu\lambda\rho} \, \tr(F^{\mu\nu} F^{\lambda\rho})
+ 16\,  P_{\mu\nu}\,  \tr(F^{\mu\lambda} F^\nu{}_{\lambda})
+ 4\, {\hat R} \,   \tr(F^{\mu\nu} F_{\mu\nu}) \, ,  \\
\noalign{\vskip 2pt}
\tr(Y^3) \to {}& {\hat R}^3 - 8\, \tr ( F^{\mu\nu} F_{\nu \lambda} F^\lambda{}_\mu)
- 6 \, {\hat R} \,   \tr(F^{\mu\nu} F_{\mu\nu}) \, ,  \\
\noalign{\vskip 2pt}
\tr(Y \nabla^2 Y) \to {}& 2 \, {\hat R} \, \nabla^2 {\hat R} -  4\,   \tr({\hat I})
+ 16\,  P_{\mu\nu}\,  \tr(F^{\mu\lambda} F^\nu{}_{\lambda})
- 16 \, {\hat R} \,   \tr(F^{\mu\nu} F_{\mu\nu}) \, ,}[trfermion]
where on the right hand side the trace is only over gauge indices, so that ${\hat I}$ is given
by \eqref{Finv} with only gauge field contributions.
To calculate the result for $\tr({\hat I})$ it is necessary to use \eqref{deriv1}
and \eqref{deriv2} to eliminate $P^{\mu\nu} \nabla^2 P_{\mu\nu}$ with
\eqna{
16 (  P^{\mu\nu} \nabla^2 P_{\mu\nu} -  {\hat R} \, \nabla^2 {\hat R} ) \to {}&
- \tfrac13(4I_1-I_2-I_3)
-8  \, P_{\mu\nu} W^{\mu\lambda \rho \omega} W^\nu{}_{\lambda\rho\omega}
+2 \, {\hat R} \,  W^{\mu\nu \lambda \rho} W_{\mu\nu\lambda\rho} \\
& + 16\, P_{\mu\nu} P_{\lambda\rho} W^{\mu\lambda\rho\nu}
+ 96\, P^{\mu\nu}P_{\nu\lambda}P^{\lambda}{}_\mu
- 16 \, {\hat R} \,  P^{\mu\nu}P_{\mu\nu} \, ,}[PRI]
discarding two derivative terms.
The traces in \eqref{trfermion} give, for $n_\psi$ fermions,
using from \eqref{Esix}
\eqna{6\, P_{\mu\nu} \, W^{\mu\lambda\rho\omega} W^\nu{}_{\lambda\rho\omega}
={}& 2 \,I_1 +   I_2  - \quar \, E_6 + \tfrac32\,  {\hat R}\,
W_{\lambda\rho}{}^{\mu\nu} W_{\mu\nu}{}^{\lambda\rho}
- 12 \, P_{\mu\nu}P_{\lambda\rho}\,  W^{\mu\lambda\rho\nu} \\
&{} + 24\,  P^{\mu\nu} P_{\nu\lambda}P^{\lambda}{}_\mu - 36\,
 {\hat R}\,  P^{\mu\nu} P_{\mu\nu} + 12\, {\hat R}^3 \, ,}[PWW]
the result from \eqref{LRF}
\eqna{L_6^R ={}&  n_\psi\, \tfrac{1}{7!} \big ( - \tfrac{1}{3} \, 14 \times 64 \, I_1 - 32 \, I_2
+ 40\, I_3 + \tfrac{191}{9} \, E_6 \big ) \, , \\
L_6^G ={}&  \tr \big ( \tfrac{4}{15} \, {\hat I} + \tfrac{2}{9} \,
W_{\mu\nu\lambda\rho} \,F^{\mu\nu} F^{\lambda\rho}
- \tfrac{52}{45} \,  F^{\mu\nu} F_{\nu \lambda} F^\lambda{}_\mu \big ) \, .}[]

\newsec{Two-forms}[resTwoForms]

We here summarise some of the results necessary in the calculation of
$a_{\Delta,3}|$ for two-forms in \eqref{L6two}.
\eqna{
& \tr_{\Omega^{(1)}} ({\hat I}) = \quar \, \tr_{\Omega^{(2)}} ({\hat I} )  \\
& \to - I_3 - 16 \,  P^{\mu\nu} \nabla^2 P_{\mu\nu} - 4\, {\hat R} \, \nabla^2 {\hat R}
+ 12\, P_{\mu\nu} W^{\mu\lambda \rho \omega} W^\nu{}_{\lambda\rho\omega}
-4 \, {\hat R} \, W^{\mu\nu\lambda\rho}W_{\mu\nu\lambda\rho}  \\
&\hskip 1cm {} + 16 \,  P_{\mu\nu} P_{\lambda\rho} W^{\mu\lambda\rho\nu}
-16 \, P^{\mu\nu}P_{\nu\lambda}P^{\lambda}{}_\mu + 40\,  {\hat R} \,
P^{\mu\nu} P_{\mu\nu} + 16 \,  {\hat R}^3  \, , \\
\noalign{\vskip 2pt}
& \tr_{\Omega^{(1)}} ( F^{\mu\nu} F_{\nu \lambda} F^\lambda{}_\mu) =
\quar \, \tr_{\Omega^{(2)}} ( F^{\mu\nu} F_{\nu \lambda} F^\lambda{}_\mu)  \\
& = - I_1 - 3 \, P_{\mu\nu} W^{\mu\lambda \rho \omega} W^\nu{}_{\lambda\rho\omega}
- 6 \, P_{\mu\nu} P_{\lambda\rho} W^{\mu\lambda\rho\nu}
- 20 \, P^{\mu\nu}P_{\nu\lambda}P^{\lambda}{}_\mu + 18 \, {\hat R} \,
P^{\mu\nu} P_{\mu\nu} + 2 \, {\hat R}^3 \, ,  \\
\noalign{\vskip 2pt}
& W_{\mu\nu\lambda\rho} \, \tr_{\Omega^{(1)}}(F^{\mu\nu} F^{\lambda\rho})
= \quar\, W_{\mu\nu\lambda\rho} \, \tr_{\Omega^{(2)}}(F^{\mu\nu} F^{\lambda\rho})  \\
&\hskip 3.6cm {}=   - I_2 -  8  \,
P_{\mu\nu} W^{\mu\lambda \rho \omega} W^\nu{}_{\lambda\rho\omega}
+ 8\, P_{\mu\nu} P_{\lambda\rho} W^{\mu\lambda\rho\nu} \, ,  \\
\noalign{\vskip 2pt}
& \tr_{\Omega^{(1)}} ( F^{\mu\nu}F_{\mu\nu} \, Y_1 ) =
- 4\,P_{\mu\nu} W^{\mu\lambda \rho \omega} W^\nu{}_{\lambda\rho\omega}
+ {\hat R}\,W^{\mu\nu\lambda\rho}W_{\mu\nu\lambda\rho}
+ 16\, P_{\mu\nu} P_{\lambda\rho} W^{\mu\lambda\rho\nu}  \\
&\hskip 3.4cm - 16\,  P^{\mu\nu}P_{\nu\lambda}P^{\lambda}{}_\mu
-8 \, {\hat R}\,  P^{\mu\nu}P_{\mu\nu}  + 4 \, {\hat R}^3   \\
& \hskip 3.4cm{} - U_{\mu\nu} W^{\mu\lambda \rho \omega} W^\nu{}_{\lambda\rho\omega}
- U \, W^{\mu\nu\lambda \rho} W_{\mu\nu\lambda\rho}
+ 4\, U_{\mu\nu} P_{\lambda\rho} W^{\mu\lambda\rho\nu}   \\
&\hskip 3.4cm {}- 4\,  P_\mu {\!}^\nu P_\nu{\!}^\rho \, U_\rho{\!}^\mu
- 2\, P_\mu {\!}^\nu P_\nu{\!}^\mu ( U_\rho{\!}^\rho + 8 \,  U  )
- 4 \, {\hat R} \, P_\mu{\!}^\nu U_\nu{\!}^\mu  - 4   \,  {\hat R}^2 \, U \, ,  \\
\noalign{\vskip 2pt}
& \tr_{\Omega^{(2)}} ( F^{\mu\nu}F_{\mu\nu} \, Y_2 ) = - I_2
- 12\,P_{\mu\nu} W^{\mu\lambda \rho \omega} W^\nu{}_{\lambda\rho\omega}
- 2\, {\hat R}\,W^{\mu\nu\lambda\rho}W_{\mu\nu\lambda\rho}
+ 24\, P_{\mu\nu} P_{\lambda\rho} W^{\mu\lambda\rho\nu}  \\
&\hskip 3.4cm
- 16\,  P^{\mu\nu}P_{\nu\lambda}P^{\lambda}{}_\mu
- 56 \, {\hat R}\,  P^{\mu\nu}P_{\mu\nu}  - 8\, {\hat R}^3  \\
& \hskip 3.4cm{} - 2\, U_{\mu\nu} W^{\mu\lambda \rho \omega} W^\nu{}_{\lambda\rho\omega}
- (U_\omega{\!}^\omega + 4\, U) \, W^{\mu\nu\lambda \rho} W_{\mu\nu\lambda\rho}
+ 8\, U_{\mu\nu} P_{\lambda\rho} W^{\mu\lambda\rho\nu}   \\
&\hskip 3.4cm {}- 8\,  P_\mu {\!}^\nu P_\nu{\!}^\rho \, U_\rho{\!}^\mu
- P_\mu {\!}^\nu P_\nu{\!}^\mu ( 20\, U_\rho{\!}^\rho + 64 \,  U  )
- 8 \, {\hat R} \, P_\mu{\!}^\nu U_\nu{\!}^\mu  - 4   \,  {\hat R}^2 ( U_\mu{\!}^\mu + 4\, U) \, ,  \\
\noalign{\vskip 2pt}
& \tr_{\Omega^{(0)}}(Y_0 ) = - 2\, {\hat R} + U\, ,
\ \   \tr_{\Omega^{(1)}}(Y_1 ) = - 2\, {\hat R} + 6 \, U + U_\mu{\!}^\mu\, , \ \
\tr_{\Omega^{(2)}}(Y_2 ) = 10\,  {\hat R} + 15 \, U + 5\, U_\mu{\!}^\mu \, ,  \\
\noalign{\vskip 2pt}
& \tr_{\Omega^{(0)}}(Y_0 {\!}^2) = (2 \, {\hat R} - U)^2 \, ,  \\
& \tr_{\Omega^{(1)}}(Y_1{\!}^2 ) =  16\, P^{\mu\nu}P_{\mu\nu} - 2\, {\hat R}^2
+ 8\, P_\mu{\!}^\nu U_\nu{\!}^\mu - 2\, {\hat R} \,U_\mu{\!}^\mu
- 4\, {\hat R} \, U  + U_\mu{\!}^\nu U_\nu{\!}^\mu
+ 2 \, U  U_\mu{\!}^\mu + 6\, U^2 \, ,
 \\
& \tr_{\Omega^{(2)}}(Y_2{\!}^2  ) = W^{\mu\nu\lambda\rho}W_{\mu\nu\lambda\rho} +
16\, P^{\mu\nu}P_{\mu\nu} + 4\, {\hat R} ^2 + 16\, P_\mu{\!}^\nu U_\nu{\!}^\mu
+ 4\, {\hat R} \,U_\mu{\!}^\mu + 20\, {\hat R} \, U  \\
& \hskip 2.1cm {} + 4\, U_\mu{\!}^\nu U_\nu{\!}^\mu + U_\mu{\!}^\mu \, U_\nu{\!}^\nu
+ 10\,  U  U_\mu{\!}^\mu + 15\, U^2\, .}[Dzero]
\eqna{\!\!
& \tr_{\Omega^{(0)}}(Y_0{\!}^3) = - (2 \, {\hat R} - U)^3 \, ,  \\
& \tr_{\Omega^{(1)}}(Y_1{\!}^3) = 64\, P^{\mu\nu}P_{\nu\lambda}P^{\lambda}{}_\mu
- 48 \, {\hat R} \,  P^{\mu\nu}P_{\mu\nu} + 6\, {\hat R}^3   \\
&\hskip 2.1cm {}+ 48\big ( P_\mu {\!}^\nu P_\nu{\!}^\rho \, U_\rho{\!}^\mu
+ P_\mu {\!}^\nu P_\nu{\!}^\mu \, U \big )
- 24 \, {\hat R} \, P_\mu{\!}^\nu U_\nu{\!}^\mu
+ 3 \,  {\hat R}^2 \, U_\mu{\!}^\mu  - 6 \,  {\hat R}^2 \, U   \\
&\hskip 2.1cm {} +12 \,  P_\mu{\!}^\nu \, U_\nu{\!}^\rho \, U_\rho{\!}^\mu
+  24 \, P_\mu{\!}^\nu \, U_\nu{\!}^\mu \, U - 3\, {\hat R}\,
U_\mu{\!}^\nu \, U_\nu{\!}^\mu
- 6 \, {\hat R} \big ( U_\mu{\!}^\mu \, U + U^2 \big )  \\
&\hskip 2.1cm {}  + U_\mu{\!}^\nu \, U_\nu{\!}^\rho \, U_\rho{\!}^\mu
+ 3\,  U_\mu{\!}^\nu \, U_\nu{\!}^\mu\, U + 3\,  U_\mu{\!}^\mu \, U^2
+ 6 \, U^3 \, ,  \\
& \tr_{\Omega^{(2)}}(Y_2{\!}^3) =  - I_2
+  12  \, P_{\mu\nu} W^{\mu\lambda \rho \omega} W^\nu{}_{\lambda\rho\omega}
+24\, P_{\mu\nu} P_{\lambda\rho} W^{\mu\lambda\rho\nu}
 + 16\,  P^{\mu\nu}P_{\nu\lambda}P^{\lambda}{}_\mu
+ 24\, {\hat R} \, P^{\mu\nu} P_{\mu\nu}  \\
& \hskip 2.1cm{}+ 6 \,U_{\mu\nu} W^{\mu\lambda \rho \omega} W^\nu{}_{\lambda\rho\omega}
+ 3 \, U \, W^{\mu\nu\lambda \rho} W_{\mu\nu\lambda\rho}
+ 24 \, U_{\mu\nu} P_{\lambda\rho} W^{\mu\lambda\rho\nu}
+ 6\, U_{\mu\nu} U_{\lambda\rho} W^{\mu\lambda\rho\nu}   \\
&\hskip 2.1cm {}+ 24\,  P_\mu {\!}^\nu P_\nu{\!}^\rho \, U_\rho{\!}^\mu
+ 12\, P_\mu {\!}^\nu P_\nu{\!}^\mu ( U_\rho{\!}^\rho + 4 \,  U  )
+ 24 \, {\hat R} \, P_\mu{\!}^\nu U_\nu{\!}^\mu + 12 \,  {\hat R}^2 \, U   \\
&\hskip 2.1cm {} +12 \,  P_\mu{\!}^\nu \, U_\nu{\!}^\rho \, U_\rho{\!}^\mu
+  12 \, P_\mu{\!}^\nu \, U_\nu{\!}^\mu \, ( U_\rho{\!}^\rho + 4\, U )
+ 6 \, {\hat R}\, U_\mu{\!}^\nu \, U_\nu{\!}^\mu
+12 \, {\hat R} \,  U_\mu{\!}^\mu \, U + 30\, {\hat R} \, U^2  \\
&\hskip 2.1cm {}  + 2\,  U_\mu{\!}^\nu \, U_\nu{\!}^\rho \, U_\rho{\!}^\mu
+ 3\,  U_\mu{\!}^\nu \, U_\nu{\!}^\mu ( U_\rho{\!}^\rho +  4\, U ) +
3\, U_\mu{\!}^\mu U_\nu{\!}^\nu \, U +  15 \,  U_\mu{\!}^\mu \, U^2 + 15 \,
U^3\, , \\
\noalign{\vskip 2pt}
& \tr_{\Omega^{(0)}}(Y_0 \nabla^2 Y_0) = (2\, {\hat R} - U)\, \nabla^2 (2\, {\hat R} - U)   \, ,  \\
\noalign{\vskip 2pt}
&  \tr_{\Omega^{(1)}}(Y_1 \nabla^2 Y_1)  = 16\, P^{\mu\nu} \nabla^2 P_{\mu\nu} -
 2\, {\hat R} \, \nabla^2 {\hat R}   \\
& \hskip 2.7cm {}+ 4 ( P^{\mu\nu}\, \nabla^2 U_{\mu\nu} + U^{\mu\nu}  \,
\nabla^2 P_{\mu\nu})
 - U_\mu{\!}^\mu\,  \nabla^2 {\hat R}  -  {\hat R}\,   \nabla^2U_\mu{\!}^\mu
 - 2( U \nabla^2 {\hat R} +{\hat R}\,  \nabla^2 U ) \\
& \hskip 2.7cm {} +  U^{\mu\nu}\, \nabla^2 U_{\mu\nu} + U_\mu{\!}^\mu \, \nabla^2 U
+ U\, \nabla^2 U_\mu{\!}^\mu + 6\, U \, \nabla^2 U \, ,   \\
\noalign{\vskip 2pt}
 &  \tr_{\Omega^{(2)}}(Y_2 \nabla^2 Y_2)  \to I_3  +  16\, P^{\mu\nu} \nabla^2 P_{\mu\nu} +
4\,  {\hat R} \, \nabla^2 {\hat R}
 - 16\,P_{\mu\nu} W^{\mu\lambda \rho \omega} W^\nu{}_{\lambda\rho\omega}
+ 8\, {\hat R}\,W^{\mu\nu\lambda\rho}W_{\mu\nu\lambda\rho}  \\
& \hskip 2.7cm {}+
8 ( P^{\mu\nu}\, \nabla^2 U_{\mu\nu} + U^{\mu\nu}  \, \nabla^2 P_{\mu\nu})  \\
& \hskip 2.7cm {}+
2 (U_\mu{\!}^\mu\,  \nabla^2 {\hat R}  + {\hat R}\,   \nabla^2U_\mu{\!}^\mu )
+ 10( U\, \nabla^2 {\hat R} +{\hat R}\,  \nabla^2 U ) \\
& \hskip 2.7cm {} + 4\,  U^{\mu\nu}\, \nabla^2 U_{\mu\nu} + U_\mu{\!}^\mu \nabla^2 \,
U_\nu{\!}^\nu + 5 ( U_\mu{\!}^\mu \, \nabla^2 U + U\, \nabla^2 U_\mu{\!}^\mu)
+ 15\, U \, \nabla^2 U \, .}[Done]
Combining terms as in \eqref{athree} and using \eqref{PRI}, \eqref{PWW} we
find
\eqna{ &  \tr_{\Omega^{(2)}} \big ( a_{\Delta^{(2)},3}| \big ) - 2\,
\tr_{\Omega^{(1)}} \big ( a_{\Delta^{(1)},3}| \big ) + 3\,
\tr_{\Omega^{(0)}} \big ( a_{\Delta^{(0)},3}| \big )  \\
& {}\to L_6^R  \\
& \hskip 0.4cm {} - G_6\vphantom{G}^{\mu\nu} \, U_{\mu\nu}  - \tfrac{11}{30}\,
W^{\mu\nu\lambda \rho} W_{\mu\nu\lambda\rho} \, U'   \\
\noalign{\vskip 2pt}
& \hskip 0.4cm {}
+ \big (\tfrac{32}{3} \, P^{\mu\nu} P_{\mu\nu} -12\, {\hat R}^2 \big ) U'
+ \tfrac{2}{3} ( \nabla^2 {\hat R} \, U' +  {\hat R} \, \nabla^2 U' )  \\
& \hskip 0.4cm {} - U_{\mu\nu} U_{\lambda\rho} \, W^{\mu\lambda\rho\nu}  \
+ 2\,  P^{\mu\nu} \, U_{\mu\rho} \, U_\nu{\!}^\rho  -2 \, P^{\mu\nu} U_{\mu\nu} \, U_\rho{\!}^\rho
- \tfrac{7}{3} \,  {\hat R}\, U^{\mu\nu}U_{\mu\nu}
 + \tfrac{13}{12} \, {\hat R}\, U_\mu{\!}^\mu \, U_{\nu}{\!}^\nu \\
\noalign{\vskip 2pt}
& \hskip 0.4cm {}
+ \tfrac{1}{6} \, \big (  U^{\mu\nu}\,  \nabla^2 U_{\mu\nu}
- \quar \, U_\mu{\!}^\mu \, \nabla^2 U_{\nu}{\!}^\nu \big )   \\
\noalign{\vskip 2pt}
& \hskip 0.4cm {}
 - \big  (  U^{\mu\nu}U_{\mu\nu} - \quar \, U_\mu{\!}^\mu \, U_\nu{\!}^\nu\big ) \, U'
 - 4 \, {\hat R}\, U'{}^2  -  \half\,  U'  \Delta_2 U' - U'{}^3 \, ,
}[Dtwo]
 where $L_6^R$ is given in \eqref{tsey} and
 \be
 U' = U + \half \, U_\mu{\!}^\mu \, .
 \ee
 From \eqref{defUv} $U' = \quar \, v^\mu v_\mu, \, U_{\mu\nu} = - \nabla_\mu v_\nu
 = \nabla_{\mu}\pr_\nu \ln g^2$.
 Since $G_6\vphantom{G}^{\mu\nu} \, U_{\mu\nu}  = \nabla_\mu\nabla_\nu (G_6\vphantom{G}^{\mu\nu} \ln g^2)$
 this term may be neglected and \eqref{Dtwo} leads to \eqref{L6g2} using
\eqna{
\nabla_\mu v_\nu \, \nabla_\lambda v_\rho \, W^{\mu\lambda\rho \nu} \to {}&
 -  \big  ( \half \, W^{\mu\lambda \rho \omega} W^\nu{}_{\lambda\rho\omega}-
 4 \, P_{\lambda\rho}  W^{\mu\lambda\rho \nu} - 3 \, B^{\mu\nu}\big ) \,
 v_\mu v_\nu \\
 ={}&   -  \big  ( \half \, W^{\mu\lambda \rho \omega} W^\nu{}_{\lambda\rho\omega}-
 B^{\mu\nu} +12  \, P^{\mu\lambda}P_\lambda{}^\nu
-2 \,  \nabla^2 P^{\mu\nu} + 2 \, \nabla^\mu\pr^\nu {\hat R} \big ) v_\mu v_\nu \\
&{} + 2\, P^{\lambda\rho}P_{\lambda\rho}  \, v^\mu v_\mu \, , \\
 \noalign{\vskip 2pt}
  {\hat R} \, \nabla^\mu v^\nu \, \nabla_\mu v_\nu  \to {}&  {\hat R}\,  \nabla_\mu v^\mu \,
 \nabla_\nu v^\nu - \big ( 4 \, {\hat R} \, P^{\mu\nu} + \nabla^\mu \pr^\nu {\hat R} \big )\, v_\mu v_\nu
 - \big  ( {\hat R}^2 - \nabla^2 {\hat R} \big  ) \, v^\mu v_\mu \, , \\
 \noalign{\vskip 2pt}
 P^{\mu\nu} \, \nabla_\mu v^\lambda \, \nabla_\nu v_\lambda \to {}& P^{\mu\nu} \,
 \nabla_\mu v_\nu \, \nabla_\lambda v^\lambda - \big (
 4 \, P^{\mu\lambda}P_{\lambda}{}^\nu +   {\hat R}\, P^{\mu\nu} - \half \, \nabla^2 P^{\mu\nu}
 + \nabla^\mu \pr^\nu {\hat R}  \big )\, v_\mu v_\nu \\
 \noalign{\vskip 2pt}
 & {}+  \half \, \nabla^2 {\hat R}\ v^\mu v_\mu \, , \\
 \nabla^\mu v^\nu \, \nabla^2 \nabla_\mu v_\nu \to {}& \nabla_\mu v^\mu \, \nabla^2
 \nabla_\nu v^\nu  + 12 \, P^{\mu\nu}\,  \nabla_\mu v_\nu \,  \nabla_\lambda v^\lambda
+3 \, {\hat R}  \, \nabla_\mu v^\mu\,  \nabla_\nu v^\nu \\
&{}+ \big ( - W^{\mu\lambda\rho\omega}W^\nu{}_{\lambda\rho\omega}
-60 \, P^{\mu\lambda}P_\lambda{}^\nu - 20 \, {\hat R}\, P^{\mu\nu} \\
\noalign{\vskip - 3 pt}
& \hskip 4cm {}
+ 6\,  \nabla^2 P^{\mu\nu} - 20\, \nabla^\mu\pr^\nu {\hat R} \big )\,
v_\mu v_\nu \\
&{}+ \big ( 2\, P^{\lambda\rho}P_{\lambda\rho}  - 2 \, {\hat R}^2
+ 9 \, \nabla^2 {\hat R} \big ) \, v^\mu v_\mu \, ,}[]
discarding total derivatives.

\end{appendix}

\end{document}